\documentclass[11pt]{article}
\usepackage[T1]{fontenc}
\usepackage[utf8]{inputenc}
\usepackage[english]{babel}
\usepackage{graphicx}
\usepackage{amsmath}
\usepackage{amssymb}
\usepackage{float}
\usepackage{setspace}
\usepackage{booktabs}
\usepackage{multirow}
\usepackage{geometry}
\usepackage{natbib}
\usepackage{array}
\usepackage{tablefootnote}
\usepackage{hyperref}
\usepackage{xcolor}
\usepackage{threeparttable}
\usepackage{comment}

\title{The Behavior of Inventories over the Business Cycle: \\ Evidence across Levels of Development\footnote{Contact information: Rodriguez Repeti: Instituto interdisciplinario de Economía Política de Buenos Aires - Universidad de Buenos Aires, \href{mailto:rodriguezrepeti.jm@economicas.uba.ar}{\textcolor{blue}{rodriguezrepeti.jm@economicas.uba.ar}}; Trupkin: Universidad de San Andr\'es, \href{mailto:dtrupkin@udesa.edu.ar}{\textcolor{blue}{dtrupkin@udesa.edu.ar}}.}}

\author{Juan M. Rodriguez Repeti \hskip20pt Danilo R. Trupkin}
\date{July 2026}

\begin{document}

\maketitle

\begin{abstract}


This paper documents how inventory dynamics vary across levels of development and how they respond to real and financial shocks. Using a balanced panel of 72 small open economies over 1993–2022, spanning advanced to low-income economies, we show that inventories are strongly procyclical in advanced economies but become acyclical at lower income levels, while volatility and persistence rise as development declines. Using local projections, we estimate inventory responses to a long-run productivity shock and to a financial shock identified from the spread between U.S. corporate bond yields and Treasury yields. Productivity shocks generate temporary inventory accumulation across all groups, with smoother adjustment in advanced economies and more irregular cycles in emerging and developing economies. Financial shocks trigger decumulation followed by a rebound in higher-income groups, but a delayed, unreversed decline in low-income economies.

\vspace{.6cm}


\vspace{0.2cm}


\end{abstract}

\newpage

\section{Introduction}

The study of inventories is central to understanding macroeconomic fluctuations. Although their relative weight in output is small, their capacity to explain business cycle movements is considerable. For example, Stock \& Watson (\citeyear{STOCK19993}) show that inventory investment is procyclical and accounts for approximately one quarter of output fluctuations in the United States. This stems from the fact that inventories not only reflect expectations of future demand, but also operate as a mechanism that smooths, amplifies, and transmits the effects of economic shocks \citep{AUERNHEIMER201470}. Recent evidence for the United States emphasizes that the dominant
drivers of inventory investment vary across frequencies.
\citet{rossi2026uncovering} finds that sales forecast errors account for the largest share of inventory investment at high frequencies, supporting the buffer-stock motive, whereas shocks to expected demand and supply associated with stockout avoidance dominate inventory fluctuations at business-cycle frequencies.

The literature documenting inventory stylized facts across countries remains limited. Chikán \& Kovács (\citeyear{CHIKAN20092}) analyze inventory behavior in ten OECD countries between 1987 and 2004 and find that GDP growth is the main determinant of stock levels. They further observe that, after declining during the 1990s, inventory investment resumes an upward trend, accompanied by a reduction in aggregate volatility that makes its dynamics more predictable. Nevertheless, the authors stress that business cycles are not homogeneous and that inventory volatility differs significantly across countries. In a similar vein, Ramey \& West (\citeyear{RAMEY1999863}) study inventories in the G7, with particular attention to the United States. Their results show that inventories are procyclical and amplify economic fluctuations. They also find that adjustments to changes in sales are slow, which keeps the inventory-sales ratio relatively stable over time.

The limited evidence for emerging economies suggests that inventory dynamics may differ substantially from those observed in advanced economies, although this evidence is firm- or plant-level and country-specific rather than macro and cross-country. Silva et al. (\citeyear{silva2022}), for example, show for Brazil that inventories respond differently depending on the nature of expectation shocks: they decline following positive expectations about firm-specific conditions, but increase when firms anticipate stronger future demand, with idiosyncratic expectation shocks generating stronger effects than aggregate ones. In a related framework, Alessandria et al. (\citeyear{alessandria2010great}), using Chilean plant-level data, find that firms relying on imported inputs hold larger inventories and adjust them less frequently, owing to higher ordering costs and delivery lags, suggesting that inventories may play an important role in the propagation of external and financial shocks in developing economies.


Taken together, these studies indicate that inventory behavior may depend on the degree of financial development, external exposure, and production structure. However, the existing evidence remains fragmented and largely country-specific. Most of the literature continues to focus on advanced economies, and there is still no systematic cross-country analysis comparing inventory dynamics and their responses to shocks across levels of development. As a result, the role of inventories as an adjustment mechanism under different macroeconomic and financial environments remains insufficiently understood.

This paper contributes to the literature in two ways. First, it documents a new set of cross-country stylized facts on inventory dynamics across advanced, emerging, and developing economies over the 1993--2022 period. The results show that inventory behavior differs across levels of development in terms of volatility, cyclicality, and persistence. In particular, procyclicality weakens as income and financial development decline, while inventory volatility and persistence increase in lower-income economies.

Second, the paper studies whether these cross-country differences also appear in the dynamic response of inventories to real and financial disturbances. By identifying long-run productivity shocks and external financial shocks via local projections, we uncover marked cross-country heterogeneity in transmission. While productivity shocks induce temporary inventory accumulation across all income groups, financial shocks trigger a temporary drawdown and recovery in high-income economies, but cause a delayed, persistent contraction in low-income countries. Together, these empirical findings map the cross-country landscape of inventory behavior and provide new stylized facts to discipline open-economy models of inventory investment and financial frictions.


The results show that inventories constitute an important channel for the transmission and adjustment of real and financial shocks, although the nature of this mechanism depends on the degree of economic and financial development. In advanced economies, inventory adjustments following productivity and financial shocks appear consistent with efficient production smoothing and working-capital management. In emerging and developing economies, by contrast, the estimated patterns are consistent with a greater role for financial frictions, liquidity constraints, external vulnerability, and weaker productive coordination.

These findings suggest that the standard inventory-smoothing view, largely developed from evidence on advanced economies, does not generalize uniformly across countries: intertemporal production smoothing alone cannot account for inventory dynamics in developing economies, where the frictions described above play a comparatively larger role. More broadly, the results suggest that inventory behavior may itself be informative about an economy's degree of financial and productive development.


The remainder of the paper proceeds as follows. Section~\ref{sec:data} presents the data and their processing, and documents the stylized facts across country groups. Section~\ref{sec:shocks} characterizes the shocks and implements the local projections used to evaluate their effects across country groups. Section~\ref{sec:results} discusses the results. Section~\ref{sec:conclusions} concludes.

\section{Data}\label{sec:data}

We use World Bank data to analyze the stylized facts of inventories in macroeconomic cycles. We select the following variables: (i) changes in inventories at current prices ($\Delta INV_{curr}$)\footnote{Changes in
inventories are defined as the net variation in goods held in inventories during the accounting period, including additions, withdrawals, and recurrent
losses. The series are obtained from official national accounts compiled under the System of National Accounts (SNA 1993/2008) framework. Within this methodology, changes in inventories are treated as a component of gross capital formation and capture goods produced but not yet consumed or sold during the current period. The variable is expressed in current local currency units.
Nevertheless, the use of internationally harmonized accounting standards allows for broad comparability across countries.}; (ii) Gross Domestic Product
(GDP) at constant prices (Y) and current prices ($Y_{curr}$); (iii) investment\footnote{Gross Fixed Capital Formation is considered as investment.} (I); (iv) government consumption (G); (v) imports (M); (vi)
exports (X); and (vii) population.

A potential concern is that, like other national accounts aggregates, changes in inventories may partly absorb the statistical discrepancies arising from the reconciliation of the production, income, and expenditure accounts. We assess the magnitude of this issue in \ref{App:data_discussion} and find that the discrepancy is small relative to inventory dynamics for the vast majority of the sample, so it is unlikely to bias the cyclical patterns documented below.

We then classify the countries in the sample into different groups. First, we divide them into advanced or developed economies and emerging market and
developing economies, following the International Monetary Fund classification (IMF).\footnote{\href{https://www.imf.org/en/publications/weo/weo-database/2023/april/groups-and-aggregates}{https://www.imf.org/en/publications/weo/weo-database/2023/april/groups-and-aggregates}}
Since the group of emerging market and developing economies presents marked heterogeneity, we further classify these countries by income level according
to the World Bank criterion: high-income, upper-middle-income, lower-middle-income, and low-income emerging countries. To guarantee the stability of countries within the subgroups, we count the number of years in
which each country belongs to each income category during the 1993--2022 period and classify each economy in the category in which it appears most frequently (mode).

We then process the data. First, we deflate changes in inventories using each country's implicit GDP price index ($\Delta INV$) and normalize their magnitude by calculating their ratio to the one-period-lagged GDP trend at constant prices, following a procedure similar to that of \citet{uribe2017open}.\footnote{We obtain the GDP trend by applying a log-quadratic filter.}
While \citet{uribe2017open} use the contemporaneous output trend to rescale the trade balance, we use its one-period lag to reduce possible endogeneity in the
local-projection estimations. We use this normalization to avoid scale problems when calculating relative volatilities with the remaining variables.

Once we construct the ratio, we build a balanced panel including countries with complete records for all the previously mentioned variables during the 1993--2022 period. We choose this time horizon based on representativeness, since data availability for all variables is greater for a broad proportion of countries beginning in the early 1990s. We then remove countries whose GDP represented more than 2\% of world GDP. Specifically, we exclude Germany, the United States, France, Italy, Japan, and the United Kingdom from the panel.\footnote{\ref{app:large_economies} presents robustness checks that include these large economies in both the stylized facts and the estimations. Overall,
the results remain very similar to those obtained when including these economies. Therefore, the results for the full sample of advanced economies do not appear to be driven exclusively by the presence of large economies.}
We use this criterion to study small open economies, both advanced and emerging.

We also identify countries that exhibit extreme behavior because of recurrent institutional instability or because they were constituted after 1993, such as Mauritania, Slovakia, Togo, and North Macedonia. We remove these countries because their repeated extreme values would bias the behavior of the remaining economies. At the end of this process, we obtain a panel of 72 countries.

As a result, we divide the sample into four groups: (i) 24 advanced high-income economies; (ii) 14 high- and upper-middle-income emerging market and developing economies (hereafter, high-income emerging economies); (iii) 24 lower-middle-income emerging market and developing economies (hereafter, middle-income emerging economies); and (iv) 10 low-income emerging market and developing economies (hereafter, low-income emerging economies) (Table \ref{tab:clasificacion_paises}).

\begin{table}[H]
\centering
\caption{Classification of sample countries according to the IMF and the World Bank}
\label{tab:clasificacion_paises}
\small
\renewcommand{\arraystretch}{1.8} 
\begin{tabular}{>{\centering\arraybackslash}m{3cm} >{\centering\arraybackslash}m{3cm} >{\centering\arraybackslash}m{8cm}}
\toprule
IMF Classification & WB Classification & Countries \\ \midrule

Developed or advanced & High & Australia, Austria, Belgium, Canada, Korea (Rep.), Denmark, Slovenia, Estonia, Finland, Greece, Hong Kong (China), Ireland, Iceland, Israel, Luxembourg, Macao (China), Norway, New Zealand, Netherlands, Portugal, Czech Republic, Singapore, Sweden, Switzerland. \\ \midrule

\multirow{7}{=}{\centering Emerging market and developing economies} & High and Upper-Middle & Argentina, Bahamas, Belarus, Botswana, Brazil, Chile, Costa Rica, Hungary, Malaysia, Mauritius, Mexico, Russia, South Africa, Turkey. \\ \cmidrule{2-3}

 & Lower-Middle & Algeria, Bolivia, Bulgaria, Cameroon, Colombia, Congo (Rep.), Cuba, Ecuador, Egypt, El Salvador, Guatemala, Honduras, Indonesia, Iran, Kazakhstan, Morocco, Namibia, Paraguay, Peru, Dominican Republic, Romania, Thailand, Tunisia, Ukraine. \\ \cmidrule{2-3} 

 & Low & Benin, Burkina Faso, Cambodia, Comoros, India, Kenya, Mali, Pakistan, Tanzania, Uganda. \\ \bottomrule
\multicolumn{3}{l}{Source: Authors' elaboration based on IMF and World Bank classifications.}
\end{tabular}
\end{table}

After constructing the balanced panel and processing the ratio data, we apply a quadratic filter in levels. The following regression describes the filter:

\begin{equation}
    \label{eq:inv_filter}
    \frac{\Delta inv_t}{Y_{trend, t-1}} = \beta_{0,i} + \beta_{1,i} t + \beta_{2,i} t^2 + \epsilon_{i,t}
\end{equation}

where $\Delta inv_t/Y_{trend, t-1}$ represents the rescaled change in inventories (in real terms), $t$ is the time variable (expressed in years), $\beta_{0,i}$, $\beta_{1,i}$, and $\beta_{2,i}$ are the regression coefficients, and $\epsilon_{i,t}$ is the error term. The error from the previous regression represents the cyclical component of the rescaled change in inventories ($\Delta inv_t^{c}$). We then take the logarithmic first difference of the expenditure and trade aggregates, previously expressed in per capita terms. These procedures
separate the cyclical and trend components of each variable. Using the resulting series, we calculate correlations to study the cyclical behavior of
inventory changes, their serial correlation, and their relationship with the remaining macroeconomic aggregates. We also calculate standard deviations to analyze their absolute and relative volatility.

After processing the data, the median volatility of changes in inventories is lower in advanced economies than in emerging and developing economies. The median is 0.70\% in advanced economies, compared with 1.13\% in emerging high-income economies, 1.22\% in emerging middle-income economies, and 1.09\% in emerging low-income economies. Taken together, emerging economies exhibit, on average, inventory volatility approximately 1.7 times that observed in advanced economies. Thus, although differences also exist across emerging economy groups, the main empirical distinction is between advanced and emerging economies.

This gap may reflect differences in financial development, macroeconomic stability, production structure, and inventory management. Advanced economies typically combine deeper credit markets, more stable macroeconomic environments, and more sophisticated supply-chain practices. These characteristics allow firms to finance working capital, adjust inventories intertemporally, and smooth production in response to temporary shocks. This interpretation is consistent with the inventory-smoothing and supply-chain management literature \citep{irvine2005inventory, CHIKAN20092}. By contrast, firms in emerging economies are generally more exposed to liquidity constraints, external shocks, and disruptions in the supply of imported inputs, which may amplify inventory fluctuations.

\begin{figure}[H]
    \centering
    \caption{Volatility of changes in inventories by country groups}
    \label{ghp:absolute_volatility}
    \includegraphics[width=1\textwidth]{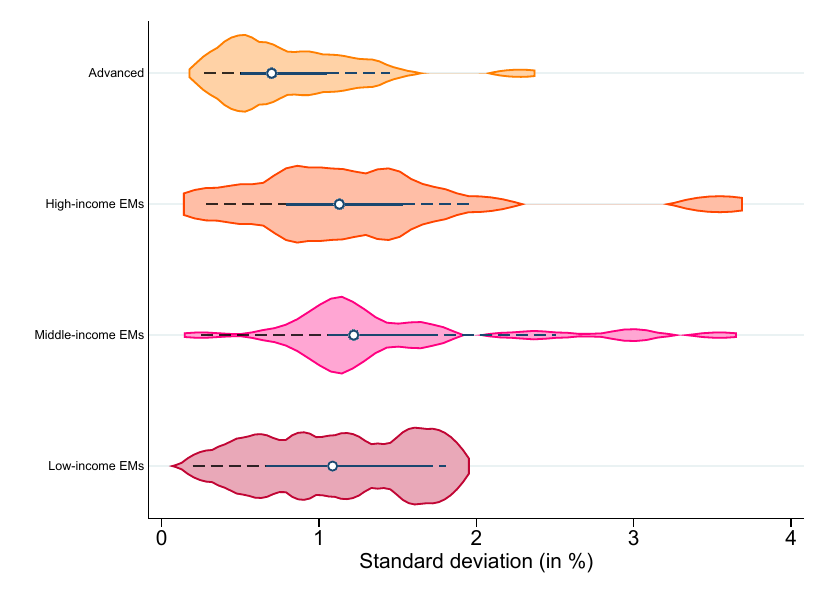}
     \begin{tabular}{p{15.0cm} p{width}}
    {\setstretch{1.1}
    \footnotesize
    \textit{Source}: Authors' elaboration based on World Bank data.}
    \end{tabular}
\end{figure}

The distributions reported in Figure \ref{ghp:absolute_volatility} provide additional evidence of these differences. Advanced economies display a relatively concentrated distribution at low levels of volatility, with most observations close to 0.6\%. Apart from a small number of exceptional observations above 2\%, this limited dispersion suggests comparatively homogeneous inventory dynamics across countries.

Emerging economies exhibit greater within-group heterogeneity. Among emerging high-income economies, the distribution is more dispersed and extends further toward high volatility values, with several observations above 3\%. 
The highest values in the upper tails of the first two country groups correspond to Singapore, classified as an advanced economy, with an inventory-cycle volatility of 2.28\%, and Botswana, classified as a high-income emerging economy, with a volatility of 3.55\%. Singapore's relatively high value may be related to its role as a major international trade and logistics hub and to the importance of electronics and semiconductor production, which are particularly exposed to fluctuations in global demand and supply chains. Botswana's high value may instead reflect the economy's strong dependence on diamond production and exports, which can generate large changes in measured inventories when external demand or the timing of production, accumulation, and sales varies.

Emerging middle-income economies present an even wider distribution and a longer right tail. Although their density is concentrated around 1\%, country-level volatility ranges from 0.25\% to 3.54\%, indicating considerable differences in exposure to shocks and in firms' ability to smooth inventory adjustment.

Emerging low-income economies also display cross-country heterogeneity, but their upper tail is more compressed and does not contain the extreme volatility levels observed in the other emerging groups. This may partly reflect the more limited role of inventories in economies characterized by lower industrial diversification, shorter production chains, and weaker integration into international trade.

The cyclical behavior of inventory changes exhibits a clear development gradient. For the full sample, inventory changes are procyclical, with a correlation with output of 0.33. They also co-move positively with the main expenditure and trade aggregates, most strongly with imports (0.33), followed by investment (0.24), private consumption (0.21), and exports (0.16). By contrast, their correlation with government consumption is weak (0.08) (Table \ref{tab:hechos_estilizados}).

This procyclicality is strongest in advanced economies and declines with income among emerging economies. The correlation between inventory changes and output is 0.48 in advanced economies and 0.39 in emerging high-income economies, but falls to 0.23 in emerging middle-income economies and becomes close to zero in emerging low-income economies ($-0.02$). A similar pattern is observed for most expenditure and trade variables. Imports display the strongest co-movement with inventories in advanced economies (0.45), emerging high-income economies (0.44), and emerging middle-income economies (0.25). In emerging low-income economies, however, inventory changes are only weakly associated with imports (0.12) and government consumption (0.10). Overall, inventory accumulation appears to be more closely connected to business-cycle and trade fluctuations in higher-income economies, whereas this relationship becomes weaker and less systematic at lower levels of development.

Inventory changes are less volatile than output and the other macroeconomic aggregates in every country group. However, their volatility relative to output increases as income declines. The ratio rises from 0.26 in advanced economies to 0.31 in emerging high-income economies, 0.37 in emerging middle-income economies, and 0.54 in emerging low-income economies. Thus, inventory fluctuations become progressively larger relative to output fluctuations across levels of development. In all groups, relative volatility is generally higher with respect to private and government consumption and considerably lower with respect to investment, exports, and imports, reflecting the greater volatility of investment and international trade flows.

Finally, the first-order autocorrelation of inventory changes is positive but moderate, reaching 0.18 for the full sample. Persistence is lowest in advanced economies (0.13), increases slightly in emerging high-income economies (0.17), and reaches 0.24 in both emerging middle-income and emerging low-income economies. This pattern is consistent with faster stock adjustment in advanced economies and greater inertia in lower-income economies, potentially reflecting differences in access to working-capital financing, supply-chain flexibility, and firms' capacity to respond to macroeconomic shocks (Table \ref{tab:hechos_estilizados})\footnote{The qualitative results are robust to alternative filtering methods. The corresponding evidence is reported in \ref{app:filtering_methods}.}.

\begin{table}[H]
    \centering
    \caption{Stylized Facts: Correlations and Relative Volatilities}
    \label{tab:hechos_estilizados}
    \small
    \renewcommand{\arraystretch}{1.1}

    \begin{tabular}{lccccc}
        \toprule

        \multicolumn{1}{c}{
            \multirow[c]{3}{*}{\textbf{Indicator}}
        }
        &
        \multirow[c]{3}{*}{
            \shortstack[c]{\textbf{Total}\\[2pt]\textbf{(a)}}
        }
        &
        \multirow[c]{3}{*}{
            \shortstack[c]{\textbf{Advanced}\\[2pt]\textbf{(b)}}
        }
        &
        \multicolumn{3}{c}{\textbf{Emerging}} \\

        \cmidrule(lr){4-6}

        & &
        & \textbf{High-income}
        & \textbf{Middle-income}
        & \textbf{Low-income} \\

        & &
        & \textbf{(c)}
        & \textbf{(d)}
        & \textbf{(e)} \\

        \midrule

        \multicolumn{6}{l}{\textit{Correlations}} \\

        Corr ($\Delta inv_t^{c}$; $\Delta y$)
        & 0.33 & 0.48 & 0.39 & 0.23 & -0.02 \\

        Corr ($\Delta inv_t^{c}$; $\Delta c$)
        & 0.21 & 0.37 & 0.38 & 0.17 & -0.06 \\

        Corr ($\Delta inv_t^{c}$; $\Delta i$)
        & 0.24 & 0.32 & 0.28 & 0.19 & -0.05 \\

        Corr ($\Delta inv_t^{c}$; $\Delta g$)
        & 0.08 & 0.09 & 0.05 & -0.04 & 0.10 \\

        Corr ($\Delta inv_t^{c}$; $\Delta x$)
        & 0.16 & 0.33 & 0.23 & 0.05 & -0.02 \\

        Corr ($\Delta inv_t^{c}$; $\Delta m$)
        & 0.33 & 0.45 & 0.44 & 0.25 & 0.12 \\

        \midrule

        \multicolumn{6}{l}{\textit{Relative volatility}} \\

        $\sigma(\Delta inv_t^{c})/\sigma(\Delta y)$
        & 0.30 & 0.26 & 0.31 & 0.37 & 0.54 \\

        $\sigma(\Delta inv_t^{c})/\sigma(\Delta c)$
        & 0.32 & 0.33 & 0.33 & 0.32 & 0.29 \\

        $\sigma(\Delta inv_t^{c})/\sigma(\Delta i)$
        & 0.11 & 0.11 & 0.11 & 0.12 & 0.12 \\

        $\sigma(\Delta inv_t^{c})/\sigma(\Delta g)$
        & 0.30 & 0.37 & 0.39 & 0.27 & 0.13 \\

        $\sigma(\Delta inv_t^{c})/\sigma(\Delta x)$
        & 0.14 & 0.11 & 0.15 & 0.15 & 0.10 \\

        $\sigma(\Delta inv_t^{c})/\sigma(\Delta m)$
        & 0.12 & 0.12 & 0.09 & 0.13 & 0.10 \\

        \midrule

        \multicolumn{6}{l}{\textit{Serial correlations}} \\

        Corr ($\Delta inv_t^{c}$; $\Delta inv_{t-1}^{c}$)
        & 0.18 & 0.13 & 0.17 & 0.24 & 0.24 \\

        \bottomrule
    \end{tabular}

    \vspace{2pt}

    {\raggedright
    \footnotesize
    Source: Authors' calculations based on World Bank data.
    \par}
\end{table}

In summary, the analysis identifies several stylized facts. First, the absolute volatility of inventory changes is lower in advanced economies than in emerging and developing economies. On average, emerging economies exhibit approximately 1.7 times the inventory volatility observed in advanced economies. Advanced economies also display a more concentrated distribution, whereas emerging economies show greater cross-country dispersion and a more frequent occurrence of high-volatility observations. These differences may reflect disparities in financial depth, macroeconomic stability, production structure, and firms' capacity to manage inventories and absorb temporary shocks.

Second, the procyclicality of inventory changes weakens as income and development decline. The correlation with output is highest in advanced economies, remains positive in emerging high-income and emerging middle-income economies, and becomes negligible in emerging low-income economies. A broadly similar pattern is observed in the co-movement of inventories with private consumption, investment, exports, and imports. This suggests that inventory accumulation is more closely linked to business-cycle fluctuations in higher-income economies, while inventories may play a more limited role in production smoothing in less developed economies. Government consumption, by contrast, displays only a weak and unsystematic relationship with inventory changes.

Third, imports exhibit the strongest and most robust association with inventory changes among the expenditure and trade aggregates considered. Their correlation is positive in every country group, although it declines at lower income levels. This pattern is consistent with inventories containing imported intermediate inputs and final goods, particularly in economies that are more integrated into international production and trade networks.

Fourth, inventory changes are less volatile than output and the other macroeconomic aggregates in all country groups. However, their volatility relative to output increases as income declines, rising from advanced economies to emerging low-income economies. Inventory volatility is also relatively higher with respect to private and government consumption than with respect to investment and international trade flows, which are themselves considerably more volatile.

Finally, the persistence of inventory changes increases as income and development decline. First-order serial correlation is lowest in advanced economies, somewhat higher in emerging high-income economies, and highest in emerging middle-income and emerging low-income economies. This pattern is consistent with faster inventory adjustment in advanced economies and greater inertia in lower-income economies, potentially reflecting differences in access to working-capital finance, supply-chain flexibility, and firms' capacity to respond to macroeconomic shocks.

\section{Real and Financial Shocks}\label{sec:shocks}

After documenting the stylized facts for the different country groups, we study the impact of real and financial shocks on inventory dynamics. To this end, we identify two shocks: (i) a long-run productivity shock; and (ii) a financial shock associated with global credit conditions, approximated by the spread between U.S. corporate bonds with medium credit risk (Baa) and U.S. Treasury bonds.

\subsection{Long-run productivity shock}

The identification strategy for the productivity shock follows the long-run restriction approach proposed by \citet{king1991stochastic} (KPSW). Under the balanced growth assumption, output, consumption, and investment share a common stochastic trend, while the consumption--output and investment--output ratios are stationary in the long run. Following the KPSW identifying logic, innovations to this permanent component are interpreted as long-run productivity shocks, under the assumption that productivity is the only source of permanent fluctuations in real aggregates.

To identify this shock, we define the vector $X_{i,t} = \begin{bmatrix} y_{i,t}^g & c_{i,t} & i_{i,t} \end{bmatrix}^{\prime}$ for each country ($i$), where $y_{i,t}^g$ denotes output excluding government consumption\footnote{A potential concern is that output includes changes in inventories as a component of expenditure-side GDP. This mechanical link is unlikely to drive the identified permanent productivity shock for three reasons. First, the median absolute ratio of changes in inventories to GDP in the sample is below 1 percent, limiting the scope for mechanical contamination. Second, the KPSW shock is recovered from the permanent component of the joint system through $C(1)$, rather than from contemporaneous inventory fluctuations. Third, changes in inventories are predominantly transitory, so their contribution to the identified permanent component is expected to be limited.}, $c_{i,t}$ denotes private consumption, and $i_{i,t}$ denotes investment. All variables are expressed in per capita terms and logarithms. We then impose a cointegration rank equal to two, with the cointegrating vectors corresponding to the balanced-growth relationships:

\begin{equation}
    \beta' X_{i,t} = 
    \begin{bmatrix} -1 & 1 & 0 \\ -1 & 0 & 1 \end{bmatrix} 
    \begin{bmatrix} y^g_{i,t} \\ c_{i,t} \\ i_{i,t} \end{bmatrix} = 
    \begin{bmatrix} c_{i,t} - y^g_{i,t} \\ i_{i,t} - y^g_{i,t} \end{bmatrix}
\end{equation}

These cointegrating vectors constitute the economically substantive content of the KPSW identification: they define the common stochastic trend implied by balanced growth. Under the maintained KPSW identifying assumption, innovations to this permanent component are interpreted as productivity shocks. Accordingly, we impose this restriction based on balanced-growth theory rather than estimate it from the data. It is therefore distinct from the pooling restriction imposed on the short-run dynamic parameters.

Given the short time dimension of the panel, estimating a fully unrestricted country-specific VECM would lead to unstable estimates of the long-run impact matrix. Therefore, we combine the long-run identifying restrictions of \citet{king1991stochastic} with the panel cointegration and common stochastic trend literature, which supports the use of cross-sectional information to estimate long-run relationships and error-correction dynamics in non-stationary panels with individual heterogeneity \citep{BAI2004137, Gengenbach2016}. The cointegration space $\beta$ is imposed according to:

\begin{equation}
    \Delta X_{i,t} = \mu_i + \alpha \beta' X_{i,t-1} + \sum_{j=1}^{p} \Gamma_j \Delta X_{i,t-j} + u_{i,t}
\end{equation}

where $\mu_i$ denotes country fixed effects, $\alpha$ is the matrix of adjustment coefficients, $\Gamma_j$ captures the short-run dynamics of the system, and $u_{i,t}$ is the vector of reduced-form innovations. The cointegration space $\beta$ is imposed according to the balanced-growth restrictions. The specification allows for country-specific average levels through fixed effects, while estimating the adjustment and short-run dynamic parameters at the panel level in order to obtain a stable estimate of the long-run impact matrix.

We emphasize what this pooling does and does not restrict. The identifying assumption concerns the long-run cointegration space ($\beta$), which is imposed according to the balanced-growth restrictions. By contrast, pooling $\alpha$ and $\Gamma_j$ imposes common adjustment coefficients and common short-run dynamics across countries. This is an econometric approximation used to obtain a stable estimate of the long-run impact matrix in a short annual panel, rather than a substantive assumption that all countries share identical business-cycle dynamics. The pooled panel VECM should therefore be interpreted as a regularized estimate of the long-run mapping from reduced-form innovations to the permanent component of output.
The pooled panel VECM should therefore be interpreted 
as a regularization device for the estimation of $C(1)$, 
not as a restriction on the identifying assumption itself\footnote{The robustness of these responses to the pooling restriction and to country-specific long-history reconstructions is examined in  \ref{app:rob_productivity} and \ref{app:long_history_validation}.}.

Under this specification, the long-run impact matrix 
is given by:

\begin{equation}
    C(1) = \beta_{\perp} \left[ \alpha_{\perp}^{\prime} \left( I - \sum_{j=1}^{p} \Gamma_j \right) \beta_{\perp} \right]^{-1} \alpha_{\perp}^{\prime}
\end{equation}

where $\beta_{\perp}$ and $\alpha_{\perp}$ denote the orthogonal complements of $\beta$ and $\alpha$, respectively. We recover the long-run productivity shock as the innovation to the permanent component of output:

\begin{equation}
    \mathit{shock}^{PS}_{i,t} = e_y^{\prime} \, C(1) \, u_{i,t}
\end{equation}

where $e_y^{\prime} = \begin{bmatrix} 1 & 0 & 0 \end{bmatrix}$ selects the long-run effect on output. We normalize the sign of the shock so that a positive value corresponds to a positive long-run effect on output. Although we estimate the long-run mapping at the panel level, the recovered shock remains country-year specific because we construct it from the reduced-form innovation $u_{i,t}$, which varies freely across countries and years. In the second stage, we accommodate heterogeneity in inventory dynamics across levels of development by estimating the local projections separately for each country group.

Before estimating the restricted panel VECM, we conduct unit-root and cointegration tests on the panel. To assess the presence of unit roots in the variables in levels, we use the IPS (Im-Pesaran-Shin) test, whose null hypothesis states that the series contain a unit root. The results indicate that this hypothesis cannot be rejected for output, consumption, and investment in levels, suggesting that the variables are non-stationary. We then apply the LLC (Levin-Lin-Chu) test to the logarithmic first differences of the variables. Since the null hypothesis of this test also corresponds to the presence of a unit root, rejection of the null indicates that the variables in first differences are stationary, that is, integrated of order one.

We also evaluate the stationarity of the balanced-growth ratios $(c_t-y_t)$ and $(i_t-y_t)$ using IPS and LLC tests with automatic lag selection based on the AIC criterion, resulting in an average lag length close to one lag. The results show evidence of stationarity for both ratios, particularly robust in the case of the investment-output relationship. Complementarily, Pedroni cointegration tests reject the null hypothesis of no cointegration between output, consumption, and investment, indicating the existence of a stable long-run relationship among the variables (Table \ref{tab:tests_panel}).

Taken together, these results are consistent with the balanced growth assumption and justify the use of panel VECM to identify the productivity shock.

\begin{table}[H]
\centering
\caption{Unit Root and Cointegration Tests}
\label{tab:tests_panel}
\small
\begin{tabular}{lccc}
\toprule
\textbf{Test} & \textbf{Variable / Relationship} & \textbf{Statistic} & \textbf{p-value} \\
\midrule

\multicolumn{4}{l}{\textit{Unit root in levels}} \\

IPS & $y_t^g$ & -0.39 & 0.35 \\
IPS & $c_t$ & 3.49 & 1.00 \\
IPS & $i_t$ & -1.27 & 0.10 \\

\midrule

\multicolumn{4}{l}{\textit{Unit root in first differences}} \\

IPS & $\Delta y_t^g$ & -29.63 & 0.00 \\
IPS & $\Delta c_t$ & -26.25 & 0.00 \\
IPS & $\Delta i_t$ & -27.97 & 0.00 \\
LLC & $\Delta y_t^g$ & -27.58 & 0.00 \\
LLC & $\Delta c_t$ & -24.83 & 0.00 \\
LLC & $\Delta i_t$ & -27.81 & 0.00 \\

\midrule

\multicolumn{4}{l}{\textit{Stationarity of balanced growth ratios}} \\

IPS (AIC) & $c_t-y^g_t$ & -1.42 & 0.07 \\
IPS (AIC) & $i_t-y^g_t$ & -4.23 & 0.00 \\
LLC (AIC) & $c_t-y^g_t$ & -1.70 & 0.05 \\
LLC (AIC) & $i_t-y^g_t$ & -4.51 & 0.00 \\

\midrule

\multicolumn{4}{l}{\textit{Cointegration}} \\

Pedroni PP & $(y_t^g,c_t,i_t)$ & -5.29 & 0.00 \\
Pedroni ADF & $(y_t^g,c_t,i_t)$ & -2.26 & 0.01 \\

\bottomrule
\end{tabular}

\vspace{0.2cm}

\begin{minipage}{0.94\textwidth}
\footnotesize{
Source: Authors' calculations based on World Bank data.\\
\textit{Note:} Variables are expressed in logarithms and per capita terms. Tests on balanced growth ratios use automatic lag selection based on the AIC criterion. The evidence against the unit-root null for investment in levels is only marginal at the 10\% level, although the evidence is less conclusive than for output and consumption; the stationarity of the investment-output ratio nonetheless provides strong support for the cointegration structure.}
\end{minipage}

\end{table}

Finally, we evaluate the temporal persistence of the identified shock through an autoregressive regression with country fixed effects and clustered standard errors at the country level. The null hypothesis of this estimation establishes the absence of temporal persistence of the shock, that is, that the coefficient associated with its first lag is equal to zero. The results show that the estimated coefficient for $shock^{PS}_{t-1}$ is small and statistically insignificant ($\hat{\rho}=-0.06$, $p$-value = 0.94), so the null hypothesis cannot be rejected. This indicates that the recovered productivity innovation does not display relevant serial correlation and supports its interpretation as an unpredictable shock to the permanent component of output.

\subsection{Identification of the Financial Shock}

We also construct a shock aimed at capturing financial frictions and exogenous changes in international credit conditions. To this end, we use the BAA spread, defined as the differential between the yield on U.S. corporate bonds rated Baa and the risk-free rate on U.S. Treasury bonds. This indicator is widely used in the macro-financial literature to approximate the external risk premium, uncertainty, liquidity constraints, and global financial conditions \citep{JUVENAL2024103913}. An increase in the spread reflects a tightening of financial conditions that raises the cost of external funding.

However, since the spread responds endogenously to economic cycles, we use an identification strategy based on instrumental variables (IV). In particular, we instrument the spread using: (i) Federal Reserve monetary surprises; and (ii) an indicator of global macro-financial uncertainty, following the approach of \citet{JUVENAL2024103913}.

We identify monetary surprises through high-frequency variations in three-month federal funds futures contracts ($FF4$) around announcements of the Federal Open Market Committee \citep{ACOSTA2024111873}. Since these data are available at daily frequency, we aggregate them to annual frequency by summing the events occurring within each calendar year:

\begin{equation}
    SM_t = \sum_{j \in t} FF4_j
\end{equation}

where $SM_t$ represents monetary surprises and $j$ indexes the monetary policy events occurring in year $t$.

On the other hand, we approximate global macro-financial uncertainty using the index developed by Ludvigson et al. (\citeyear{Ludvigson2021}) \footnote{We use updated data series from the authors' website to obtain data for 2022.}. Since this indicator is available at monthly frequency, we transform it into annual frequency using the arithmetic average of its
observations within each year.

Using these variables, we estimate the BAA spread according to:

\begin{equation}
    BAA_{t} = \gamma_0 + \gamma_1 SM_t + \gamma_2 UNC_t + \nu_t
\end{equation}

where $BAA_t$ is the observed corporate risk spread in period $t$; $\gamma_0$ is the regression constant; $SM_t$ represents Federal Reserve monetary surprises; $UNC_t$ is the global macro-financial uncertainty index; and $\nu_t$ is the error term. We define the financial shock as the instrumented component of the BAA spread, that is, the variation in the spread predicted by the external instruments.

We evaluate the identification strategy for the financial shock through econometric tests of instrument relevance, instrument exogeneity, and shock persistence.

First, we evaluate instrument relevance in the first-stage regression for the BAA spread. Since the financial shock is global and varies only over time, we estimate the first stage using the annual global series rather than the repeated country-year panel. The joint significance test of Federal Reserve monetary surprises and global macro-financial uncertainty yields an $F$-statistic of $10.69$ ($p$-value = $0.0004$). This value is slightly above the conventional rule-of-thumb threshold of 10 proposed by \citet{staiger1997instrumental}, suggesting that the instruments have relevant explanatory power for the BAA spread and alleviating, although not eliminating, concerns about weak instruments.

Because conventional first-stage $F$-statistics may be insufficient in the presence of heteroskedasticity and clustered errors, we complement the diagnostic with the effective $F$-statistic proposed by \citet{montielolea2013robust}. We compute the statistic in the LP-IV specifications, partial out country fixed effects, and cluster standard errors by year. The effective $F$-statistics range between $8.83$ and $14.85$ across country groups and horizons. These values generally exceed the critical values associated with a 20 percent maximal relative bias of TSLS, although they do not uniformly exceed the more stringent 10 percent threshold. Therefore, the instruments display moderate relevance, and we interpret the
financial-shock estimates with appropriate caution.\footnote{\ref{app:specifications} reports the complete Montiel Olea--Pflueger effective $F$-statistics by country group and horizon.}

Second, we assess the overidentifying restrictions using the Hansen J test. The null hypothesis is that the instruments are orthogonal to the error term. The test does not reject this hypothesis ($p$-value = 0.58), so we fail to reject the overidentifying restrictions.

Third, we evaluate the endogeneity of the BAA spread using the Durbin-Wu-Hausman test. The test rejects the null hypothesis of exogeneity ($p$-value = 0.0038), supporting the use of an instrumental-variables strategy instead of ordinary least squares.

Finally, we evaluate the temporal persistence of the instrumented financial shock using a first-order autoregressive specification with country fixed effects. The coefficient on the lagged shock is positive but statistically insignificant ($\rho=0.22$, $p$-value = 0.737), suggesting that the identified financial shock exhibits a predominantly transitory behavior (Table \ref{tab:test_shock_spread}). Low-income EMDEs have the weakest diagnostics, with effective F-statistics ranging between $8.83$ and $13.03$. These values are above, or close to, the critical values associated with a 20 percent maximal relative bias of TSLS, but they do not uniformly exceed more stringent thresholds.

\begin{table}[H]
    \centering
    \caption{Validation Tests for the Financial Shock}
    \label{tab:test_shock_spread}
    \begin{tabular}{lcc}
    \hline
    \textbf{Test} & \textbf{Statistic} & \textbf{p-value} \\
    \hline
    First-stage $F$ (global relevance) & 10.69 & 0.000 \\
    Montiel Olea--Pflueger effective $F$ & 8.83--14.85 & -- \\
    Hansen J (overidentification) & -- & 0.58 \\
    Durbin-Wu-Hausman (endogeneity) & 8.98 & 0.0038 \\
    \hline
    \multicolumn{3}{l}{\textit{Shock persistence (AR(1) with fixed effects)}} \\
    \hline
    $\rho$ (shock$_{t-1}$) & 0.2188 & 0.737 \\
    \hline
    \end{tabular}
    \begin{flushleft}
    \footnotesize Source: Authors' calculations based on World Bank data.\\
    Notes: We compute the first-stage $F$-statistic using the annual global series, since the BAA spread and the instruments vary only over time. The Hansen J test evaluates the overidentifying restrictions, while the Durbin-Wu-Hausman test evaluates the exogeneity of the BAA spread in the structural equation.
    \end{flushleft}
\end{table}

\subsection{Local Projections}

We use the local-projection method developed by \citet{jorda} to estimate the dynamic response of inventories to structural shocks. This approach allows us
to estimate impulse-response functions flexibly through independent regressions for each projection horizon.

We conduct the analysis separately for each identified shock. In both cases, we estimate the dynamic response of the cyclical component of inventory changes ($\Delta inv_t^c$). For each horizon $h = 0,1,\dots,H$, with $H=5$, we estimate the following specifications:

\begin{equation}
\Delta inv^{c}_{i,t+h} = \alpha_i + \lambda_t + \beta^{PS}_h shock^{PS}_{i,t} + \Gamma^{PS}_h Z_{i,t-1} + \varepsilon^{PS}_{i,t+h}
\end{equation}

\begin{equation}
\Delta inv^{c}_{i,t+h} = \alpha_i + \beta^{F}_h \widehat{BAA}_{t} + \Gamma^{F}_h Z_{i,t-1} + \varepsilon^{F}_{i,t+h}
\end{equation}

where $\Delta inv_{i,t+h}^c$ represents the cyclical component of inventory changes at horizon $t+h$, and $shock^{PS}_{i,t}$ and $\widehat{BAA}_{t}$ are the productivity and financial shocks, respectively. The terms $\alpha_i$ and $\lambda_t$ represent country and year fixed effects, respectively. Since the financial shock is global and varies only over time, \ref{app:specifications} reports additional robustness exercises using year-clustered standard errors, two-way clustering by country and year, and two-way clustering combined with global crisis dummies. The qualitative pattern of the estimated responses remains stable across these alternative inference schemes. The coefficients $\beta^{PS}_h$ and $\beta^{F}_h$ measure the dynamic response of inventories to each type of shock.

The vector $Z_{i,t-1}$ contains the control variables, namely two lags of output growth and two lags of the cyclical component of inventory changes, in order to capture the dynamics of the system\footnote{\ref{app:specifications_1} presents robustness checks using alternative control specifications for each shock, and \ref{app:leave_one_country_out} presents leave-one-country-out robustness checks for each shock}. The vectors $\Gamma^{PS}_h$ and $\Gamma^{F}_h$ collect the coefficients associated with these controls at each horizon.

We estimate the specifications separately for each country group and each shock, allowing for heterogeneity in the dynamic response. Following \citet{jorda}, we estimate each horizon independently without imposing global dynamic restrictions.

Finally, we estimate the regressions using standard errors clustered at the country level to account for possible heteroskedasticity and serial dependence
within each unit. For the productivity shock, we also use a robust estimator to reduce the influence of outliers in some countries in the sample \citep{Hamilton1991, Rousseeuw1987}\footnote{Given the cross-country nature of the panel and the presence of common global shocks, \ref{app:csd_tests} reports Pesaran CD tests for residual cross-sectional dependence. The evidence motivates the robustness exercises based on alternative inference schemes, including year-clustered and two-way clustered standard errors.} .

To complement the impulse-response analysis, we compute an LP-based variance share for each shock, country group, and projection horizon. We use this statistic to quantify the fraction of residualized inventory variation accounted for by the shock-implied component of the response. We interpret it as a local-projection analogue of an incremental variance decomposition rather than as a VAR-based forecast-error variance decomposition. For each group $g$, horizon $h$, and shock $j$, we define the variance share as: 

\begin{equation}
\label{eq:variance}
VS^{j}_{g,h} = 100 \times \frac{ \widehat{\beta}^{j\,2}_{g,h} \operatorname{Var}_{w}\left(\widetilde{s}^{j}_{i,t}\right) }{ \operatorname{Var}_{w}\left(\widetilde{\Delta inv}^{c}_{i,t+h}\right) }, 
\end{equation} 

where $\widehat{\beta}^{j}_{g,h}$ is the local-projection coefficient associated with shock $j$, $\widetilde{s}^{j}_{i,t}$ denotes the shock residualized with respect to the baseline controls and fixed effects, and $\widetilde{\Delta inv}^{c}_{i,t+h}$ denotes the residualized cyclical component of inventory changes. We residualize both variables with respect to the same controls and fixed effects included in the baseline local projection. The operator $\operatorname{Var}_{w}(\widetilde{s}^{j}_{i,t})$ denotes a weighted variance. In the case of the productivity shock, the weights correspond to the final observation weights from the robust local projection. For the financial shock, we compute the statistic on the LP-IV estimation sample using the instrumented component of the BAA spread. The numerator captures the variance of the inventory component implied by the identified shock, $\widehat{\beta}^{j}_{g,h}\widetilde{s}^{j}_{i,t}$, while the denominator captures the residualized variance of inventories at the same horizon. Therefore, higher values indicate that the corresponding shock accounts for a larger fraction of inventory fluctuations conditional on the controls and fixed effects used in the baseline specification. Since we compute the statistic separately for each shock, we do not interpret the resulting shares as an exhaustive decomposition of total inventory variance, and they are not required to sum to 100 percent.

\section{Results}\label{sec:results}

\subsection{Productivity shock}

\begin{figure}[H]
    \centering
    \caption{Responses of Inventory Changes to a Long-Run Productivity Shock,
    by Country Group}
    \label{fig:irf_shock_tfp}

    \includegraphics[width=.9\textwidth]{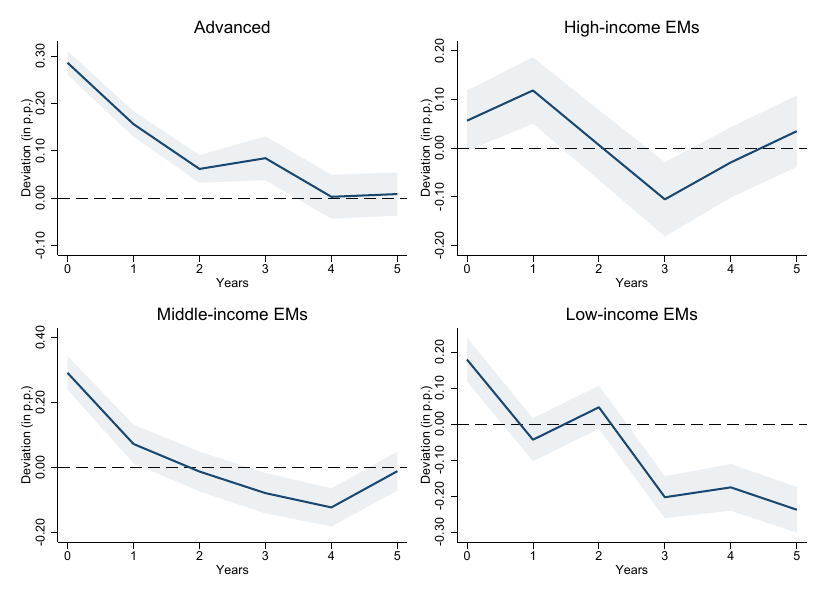}

    \vspace{0.15cm}

    \begin{minipage}{0.9\textwidth}
        \setstretch{1.1}
        \scriptsize
         \textit{Note:} Authors' calculations based on World Bank data. Shaded areas represent 68\% confidence intervals.
    \end{minipage}
\end{figure}

Figure \ref{fig:irf_shock_tfp} shows that the estimated responses of inventory changes to a positive long-run productivity shock differ across country groups in both magnitude and persistence.

In advanced economies, inventory changes rise on impact and then decline gradually toward zero. The 68\% confidence interval excludes zero from the impact period through horizon three, while the response becomes small and imprecisely estimated thereafter. The smooth convergence toward zero is
consistent with a relatively rapid adjustment of inventory holdings following the initial expansion.

In emerging high-income economies, inventory changes increase on impact and remain positive at horizon one. The response subsequently declines, becomes negative, and reaches its lowest value at horizon three before returning toward zero. The 68\% confidence interval excludes zero at horizons one and
three, whereas the remaining estimates are less precise. This pattern suggests that the initial inventory accumulation is temporary and is followed by a subsequent reduction of stocks.

Emerging middle-income economies also display an initial increase in inventory changes. The response remains positive at horizon one, approaches zero at horizon two, and becomes negative at horizons three and four before recovering
at the end of the projection horizon. The 68\% confidence interval excludes zero on impact and at horizons one, three, and four. Relative to the estimated path for advanced economies, the point estimates for emerging middle-income economies suggest a more prolonged cycle of inventory accumulation and decumulation.

In emerging low-income economies, inventory changes increase on impact, fall in the following period, recover temporarily at horizon two, and become negative from horizon three onward. The 68\% confidence interval excludes zero on impact and between horizons three and five. The alternating pattern of accumulation and decumulation indicates a less smooth adjustment process and a more persistent response at longer
horizons.

Taken together, the point estimates show two main patterns. First, the estimated adjustment path is smoother in advanced economies. Following the
initial accumulation, inventory changes decline gradually and converge toward zero without an important reversal. In emerging economies, by contrast, the initial increase is generally followed by a period of inventory decumulation.
This reversal is particularly visible in emerging middle-income and emerging low-income economies, where the response remains negative at several intermediate or longer horizons.

Second, the estimated responses tend to last longer and follow less uniform paths in emerging economies. In advanced economies, most of the response is
concentrated in the first few years and subsequently fades. Emerging high-income economies show a relatively short-lived cycle of accumulation and decumulation, while emerging middle-income economies display a longer
adjustment. Emerging low-income economies exhibit the least regular point-estimate trajectory, with alternating positive and negative responses over the projection horizon.

One possible interpretation is that firms in advanced economies may be better able to align production, sales, and desired inventory holdings after a productivity shock. Deeper working-capital markets, more developed logistics systems, and better information about demand may facilitate a smoother adjustment. In emerging economies, financial constraints, greater reliance
on imported inputs, and less flexible supply chains may delay or amplify the inventory response. These mechanisms are consistent with the inventory and supply-chain literature
\citep{kahn1987inventories,bils2000inventory,irvine2005inventory,
alessandria2013trade,sarte2015inventory}, but they should be interpreted as possible explanations rather than as mechanisms directly identified by the
local projections \footnote{\ref{app:others_variables} reports the impulse-response functions for the other macroeconomic aggregates used as auxiliary evidence
for interpreting the productivity shock.}.

The variance shares reported in Table \ref{tab:lp_variance_share_productivity} provide complementary evidence on the quantitative importance of the productivity shock and reinforce several of the patterns observed in the impulse responses. In advanced economies, the shock accounts for 9.77\% of the residualized variation in inventory changes
on impact. Its contribution then declines steadily, falling to 5.34\% at horizon one and 2.36\% at horizon two, before becoming negligible toward the end of the projection horizon. This profile is consistent with an inventory response that is concentrated at short horizons and gradually dissipates.

In emerging high-income economies, the productivity shock explains less than 1\% of the residualized variation on impact. Its contribution increases temporarily to 2.28\% at horizon one but remains below 1\% thereafter. In emerging middle-income economies, the shock accounts for 5.11\% on impact,
but its contribution falls below 1\% from horizon one onward. Thus, in both groups, the explanatory power of the productivity shock is largely concentrated in the initial periods. Although the impulse responses display subsequent reversals, these movements account for only a limited share of the overall variation in inventory changes.

Emerging low-income economies display a different pattern. The shock explains 15.78\% of the residualized variation in inventory changes on impact. Its contribution falls rapidly over the following two horizons but then rises again, reaching 14.53\% at horizon five. This irregular profile is
consistent with the more persistent and irregular adjustment observed in the impulse-response estimates.

Overall, the impulse responses and variance shares point to a common conclusion. Productivity shocks generate an initial accumulation of inventories across all country groups, but the subsequent adjustment is smoother and more concentrated at short horizons in advanced economies. The estimated responses for emerging economies display reversals, longer adjustment cycles, and less regular paths. These results suggest that inventories constitute an
important margin through which productivity shocks are absorbed, while the timing and persistence of the adjustment vary across levels of development.

\begin{table}[H]
\centering
\caption{LP-Based Variance Shares of the Long-Run Productivity Shock (in \%)}
\label{tab:lp_variance_share_productivity}
\small

\begin{tabular}{lcccccc}
\toprule
\textbf{Country group}
& \textbf{$h=0$}
& \textbf{$h=1$}
& \textbf{$h=2$}
& \textbf{$h=3$}
& \textbf{$h=4$}
& \textbf{$h=5$} \\
\midrule

Advanced economies
& 9.77 & 5.34 & 2.36 & 1.40 & 0.01 & 0.12 \\

High-income EMs
& 0.87 & 2.28 & 0.84 & 0.96 & 0.00 & 0.02 \\

Middle-income EMs
& 5.11 & 0.08 & 0.01 & 0.14 & 0.22 & 0.43 \\

Low-income EMs
& 15.78 & 3.73 & 0.65 & 2.29 & 3.92 & 14.53 \\

\bottomrule
\end{tabular}

\vspace{0.15cm}

\begin{minipage}{0.95\textwidth}
\footnotesize
\textit{Notes:} The table reports the percentage of the residualized
variation in inventory changes accounted for by the long-run productivity
shock at each projection horizon. Source: Author's elaboration based on
World Bank data.
\end{minipage}

\end{table}

\subsection{Financial shock}

\begin{figure}[H]
    \centering
    \caption{Responses of Inventory Changes to a Financial Shock, by Country Group}
    \label{fig:irf_shock_spread}
    \includegraphics[width=.9\textwidth]{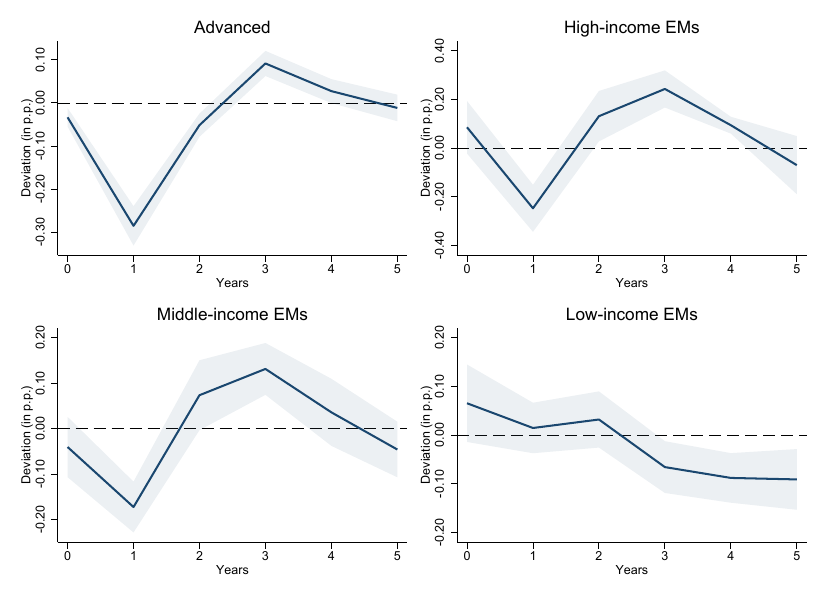}
        \begin{tabular}{p{15.0cm} p{width}}
    {\setstretch{1.1}
    \scriptsize
    \textit{Note}: Authors' elaboration based on World Bank data. Shaded areas represent 68\% confidence intervals.}
\end{tabular}
\end{figure}

Figure \ref{fig:irf_shock_spread} reports the response of inventory changes to an exogenous tightening in international financial conditions. The shock corresponds to a positive innovation in the financial component of the BAA spread identified through the LP-IV strategy described in the empirical
methodology. Economically, the shock captures a deterioration in external credit conditions that may affect inventory decisions by raising financing costs, tightening firms' liquidity constraints, and reducing their capacity to finance working capital. The estimated response paths differ across country groups in both the timing and persistence of the inventory adjustment.

In advanced economies, the contemporaneous response is small and close to zero. Inventory changes then fall quickly at horizon one and remain negative, although considerably smaller in magnitude, at horizon two. The response reverses at horizon three, when inventory accumulation becomes positive, and
subsequently converges toward zero. The 68\% confidence interval excludes zero at horizons one, two, and three.

This trajectory suggests that firms initially respond to tighter financial conditions by drawing down inventories. Such an adjustment is consistent with inventories acting as a source of liquidity when the cost of financing working capital increases. Rather than maintaining previously planned stock levels, firms may reduce purchases of inputs and use existing inventories to
support production and sales. The subsequent positive response is consistent with a rebuilding of stocks once the initial financial contraction has been absorbed. The relatively smooth return toward zero suggests that the inventory cycle is concentrated in the first few years following the shock.

Emerging high-income economies show a similar cycle, with larger point estimates at several horizons. Inventory changes are positive on impact, fall quickly at horizon one, and then recover strongly. The response becomes positive at horizon two, reaches its maximum at horizon three, remains positive at horizon four, and turns slightly negative at the end of the projection horizon. The 68\% confidence interval excludes zero for most of the responses between impact and horizon four, while the estimate at horizon five is less precise.

The large fall at horizon one indicates that inventory decumulation is an important short-run response to deteriorating external financial conditions in this group. Firms may reduce stocks to preserve liquidity, limit working-capital requirements, or compensate for more restricted access to
external financing. The point estimates show a larger rebound that remains positive for more horizons than in advanced economies. This rebound may reflect the need to restore inventories after a relatively large initial drawdown, particularly when production relies on imported intermediate inputs or when firms face uncertainty about the future availability and cost of external finance. The estimates therefore point to a marked cycle of decumulation and rebuilding, rather than to a permanent reduction in inventory holdings.

In emerging middle-income economies, the response is negative on impact but relatively small and imprecisely estimated. Inventory changes decline more strongly at horizon one, where the 68\% confidence interval excludes zero. The response then becomes positive at horizons two and three, reaching its maximum at horizon three, before declining and returning to negative values at horizon five.

The behavior again suggests that firms initially use inventory decumulation as an adjustment margin following the financial tightening. The subsequent reversal is economically visible but is surrounded by wider confidence intervals than in emerging high-income economies. This lower precision may reflect the greater heterogeneity of countries included in the group, as well as differences in their exposure to international capital markets, domestic credit conditions, trade structure, and production networks. Consequently, the international financial shock may be transmitted strongly in some countries but only weakly in others, producing a less uniform group-level response.

Emerging low-income economies display a different adjustment pattern. Inventory changes are slightly positive from impact through horizon two, but these initial responses are small and their 68\% confidence intervals include zero. From horizon three onward, the response becomes negative and remains below zero through the end of the projection horizon. The confidence interval excludes zero at horizons three, four, and five, indicating that the negative response emerges with a delay and becomes more persistent at longer horizons.

This delayed response contrasts with the immediate inventory decumulation observed in the other country groups. One possible explanation is that low-income economies are less directly integrated into international capital markets, weakening the contemporaneous transmission of changes in global financial conditions. However, tighter external finance may still affect
these economies indirectly and with a lag through trade, imported input prices, foreign exchange availability, domestic credit conditions, or lower external demand. Once these indirect effects materialize, firms may reduce inventories for several consecutive periods. The results therefore do not indicate an absence of financial transmission, but rather a delayed negative response that remains below zero through the end of the projection horizon.

Taken together, the point estimates show two main patterns across advanced and emerging economies. First, the timing of the initial inventory adjustment varies across groups. Advanced, emerging high-income, and emerging middle-income economies display a substantial inventory decumulation at
horizon one. In emerging low-income economies, by contrast, the negative response appears only from horizon three onward. This timing is consistent with a more immediate transmission of global financial conditions in economies that are more closely integrated into international financial and production networks.

Second, the point estimates show a larger subsequent rebound in the higher-income groups. Advanced economies rebuild inventories at horizon three and then converge rapidly toward zero. Emerging high-income economies exhibit an even larger rebound between horizons two and four, while the recovery in emerging middle-income economies is more moderate and less precisely estimated.
Emerging low-income economies do not exhibit a comparable rebuilding phase within the projection horizon. Instead, their response becomes increasingly negative at longer horizons.

These patterns are consistent with inventories serving both as a liquidity buffer and as a margin of production adjustment. When financing conditions tighten, firms can preserve cash by reducing purchases and meeting part of current sales from previously accumulated stocks. As financial conditions stabilize, they may subsequently rebuild inventories to restore desired stock levels. The speed and magnitude of this process are likely to depend on firms' access to working-capital finance, their reliance on imported inputs, the flexibility of their supply chains, and their ability to coordinate production and sales. This interpretation is consistent with evidence that
inventory adjustment depends on firms' liquidity, operational capacity, and working-capital management, particularly during periods of financial stress \citep{udenio2018inventory}
\footnote{\ref{app:others_variables} reports the impulse-response functions for the other macroeconomic aggregates used as auxiliary evidence for interpreting the financial shock.}.

The variance shares reported in Table \ref{tab:lp_variance_share_financial} provide complementary evidence on the quantitative importance of the financial shock.  These measures capture the horizon-specific share of residualized inventory variation associated with the estimated LP-IV response and should not be interpreted as a conventional forecast-error variance decomposition.

In advanced economies, the financial shock explains only 0.06\% of the residualized variation in inventory changes on impact. Its contribution rises sharply to 7.64\% at horizon one, before falling to 0.31\% at horizon two and remaining below 1\% throughout the rest of the projection horizon. This profile closely matches the impulse-response estimates where the shock has little contemporaneous effect but generates a substantial inventory decumulation one year later, after which its quantitative importance declines rapidly.

The concentration of the variance share at horizon one indicates that the financial shock is particularly relevant for the immediate post-shock adjustment in advanced economies. At subsequent horizons, other macroeconomic disturbances and country-specific factors account for most of the remaining variation in inventory changes. Thus, although the shock
generates a clearly identifiable inventory cycle, its explanatory
contribution is short-lived.

In emerging high-income economies, the shock explains 0.38\% of residualized inventory variation on impact and 1.99\% at horizon one. Its contribution falls to 0.80\% at horizon two, increases again to a maximum of 2.12\% at horizon three, and subsequently declines. This profile shows the decumulation-and-rebuilding cycle observed in the impulse responses where the first increase coincides with the inventory reduction at horizon one, whereas the second coincides with the subsequent rebuilding of stocks. Nevertheless, even at its maximum, the shock explains only a modest proportion of overall inventory variation.

In emerging middle-income economies, the financial shock accounts for 0.03\% of residualized inventory variation on impact and remains below 1\% at every horizon. Its largest contributions are observed at horizon one (0.77\%) and
horizon three (0.56\%), coinciding with the main negative and positive movements in the impulse response. The limited variance shares and the relatively wide confidence intervals suggest that inventory dynamics in this group are shaped by a broader range of domestic and external disturbances, reducing the explanatory importance of the common financial shock.

Emerging low-income economies also display small variance shares at short horizons. The shock explains 0.25\% of residualized inventory variation on impact, 0.02\% at horizon one, and 0.04\% at horizon two. Its contribution then rises gradually, reaching 0.54\% at horizon three, 0.89\% at horizon four, and 1.20\% at horizon five. This delayed increase is consistent with the impulse-response estimates, in which the negative inventory response emerges only from horizon three onward.

The variance-share profile therefore reinforces the interpretation of a lagged transmission mechanism in emerging low-income economies. Global financial conditions explain little of the immediate variation in inventories but become somewhat more relevant at longer horizons, as indirect effects
through trade, exchange rates, external demand, or domestic financing conditions begin to materialize. Even at horizon five, however, the financial shock continues to account for only a limited share of total residualized inventory variation.

Overall, the impulse responses and variance shares provide a coherent picture of the inventory response to an international financial tightening. In advanced economies, the effect is concentrated at horizon one and accounts for a comparatively large share of inventory variation in that period.
Emerging high-income economies display a more extended cycle, in which the shock contributes both to the initial decumulation and to the subsequent rebuilding of stocks. In emerging middle-income economies, the estimated cycle is visible but explains only a small share of overall inventory variation, whereas in emerging low-income economies the effect is initially weak but becomes more persistent at longer horizons.

These findings indicate that inventories are an economically relevant margin of adjustment to financial shocks, although their quantitative importance and timing vary across levels of development. Greater integration into global financial markets appears to be associated with a faster initial response,
while lower-income economies display a more delayed adjustment. At the same time, the generally modest variance shares outside horizon one in advanced economies show that international financial shocks are only one of several forces driving inventory fluctuations.

\begin{table}[H]
\centering
\caption{LP-IV-Based Variance Shares of the Financial Shock}
\label{tab:lp_variance_share_financial}
\small
\begin{tabular}{lcccccc}
\toprule
\textbf{Group} & \textbf{$h=0$} & \textbf{$h=1$} & \textbf{$h=2$} & \textbf{$h=3$} & \textbf{$h=4$} & \textbf{$h=5$} \\
\midrule
Advanced             & 0.06 & 7.64 & 0.31 & 0.60 & 0.06 & 0.00 \\
High-income EMs & 0.38 & 1.99 & 0.80 & 2.12 & 0.18 & 0.77 \\
Middle-income EMs         & 0.03 & 0.77 & 0.20 & 0.56 & 0.04 & 0.09 \\
Low-income EMs                 & 0.25 & 0.02 & 0.04 & 0.54 & 0.89 & 1.20 \\
\bottomrule
\end{tabular}
\begin{flushleft}
\footnotesize \textit{Notes:}  Authors' elaboration based on World Bank data.
\end{flushleft}
\end{table}

\section{Conclusions}\label{sec:conclusions}

This paper studies the cyclical properties of inventory changes and their responses to real and financial shocks in a balanced panel of 72 small open economies over the period 1993--2022. By excluding the largest economies and
comparing advanced economies with emerging market and developing economies at different income levels, the analysis focuses on countries whose business
cycles are particularly exposed to international trade and financial conditions. The main contribution is to document that inventory dynamics differ across levels of development and that the patterns commonly observed in advanced economies do not extend uniformly to emerging economies.

The descriptive evidence reveals a clear distinction between advanced and emerging economies. The median volatility of inventory changes is 0.70\% in advanced economies, compared with values between 1.09\% and 1.22\% across the three emerging economy groups. Taken together, emerging economies exhibit, on average, approximately 1.7 times the inventory volatility observed in advanced economies. Their distributions are also more dispersed, indicating greater cross-country heterogeneity in inventory dynamics.

Inventory cyclicality and persistence also vary with development. The correlation between inventory changes and output is 0.48 in advanced economies and declines progressively across emerging economy groups, becoming negligible in emerging low-income economies. Conversely, first-order serial correlation is lowest in advanced economies and highest in emerging middle-income and
emerging low-income economies. Inventory adjustment therefore appears to be more closely connected to the business cycle and less persistent in advanced economies, while lower-income economies display weaker procyclicality and greater inertia.

The relationship with international trade is particularly informative for small open economies. Imports exhibit the strongest and most systematic co-movement with inventory changes among the expenditure and trade aggregates
considered. The correlation is positive in every country group, although it declines with income. This result is consistent with inventories containing imported intermediate inputs and final goods, and suggests that trade integration is an important dimension of inventory adjustment. At the same time, the weaker relationship observed at lower income levels may reflect
shorter production chains, lower industrial diversification, and more limited participation in international production networks.

The local-projection results show that the estimated responses to long-run productivity shocks also follow different paths across country groups. A positive shock generates an initial accumulation of inventories in
all country groups. In advanced economies, the response subsequently declines smoothly and converges toward zero. Emerging economies display less uniform adjustment paths, including stronger reversals and more persistent cycles of
accumulation and decumulation. The pattern is particularly irregular in emerging low-income economies, where the initial increase is followed by a persistent contraction at longer horizons.

The LP-based variance shares provide complementary evidence on the quantitative importance of the productivity shock. In advanced economies, the shock explains almost 10\% of residualized inventory variation on impact, but its contribution declines rapidly thereafter. Its impact contribution is also sizeable in
emerging middle-income and emerging low-income economies, although the subsequent profiles differ considerably. In emerging low-income economies, the variance share rises again at longer horizons, consistent with the more
persistent response observed in the impulse-response estimates. This long-horizon result should nevertheless be interpreted cautiously given the smaller number of countries in this group.

The estimated response to a tightening in global financial conditions also follows different paths across levels of development. Advanced, emerging high-income, and emerging
middle-income economies display a substantial inventory decumulation one year after the shock, followed by a rebuilding of stocks. This pattern is consistent
with firms using inventories as a liquidity and production-adjustment margin when tighter credit conditions raise working-capital costs. In emerging
low-income economies, by contrast, the negative response appears only after several years and remains persistent through the end of the projection horizon. This timing is consistent with faster transmission in economies with stronger integration into international capital markets and with delayed effects in less financially integrated economies.

The corresponding variance shares indicate that the quantitative contribution of the financial shock is generally modest and more concentrated at specific
horizons than that of the productivity shock. In advanced economies, it explains 7.64\% of residualized inventory variation at horizon one but becomes negligible thereafter. The shares are smaller in emerging economies, although their timing is consistent with the estimated impulse responses. The contribution is associated with both inventory decumulation and subsequent
rebuilding in emerging high-income economies, remains limited at all horizons in emerging middle-income economies, and increases gradually at longer horizons in emerging low-income economies.

Taken together, these results indicate that inventories perform two related functions in small open economies. They allow firms to adjust production and sales following real shocks, and they provide a margin through which firms
respond to changes in liquidity and external financing conditions. The relative importance and timing of these functions, however, depend on the economic environment. Deeper credit markets, more developed supply chains,
and stronger production coordination may facilitate smoother and faster inventory adjustment. Financial constraints, import dependence, and greater exposure to external disruptions may instead produce more persistent or less regular responses. These mechanisms are consistent with the evidence but are not separately identified by the empirical strategy.

The findings therefore qualify the standard inventory-smoothing framework, which is based largely on evidence from advanced economies. In emerging and developing small open economies, inventory decisions appear to reflect not
only intertemporal production smoothing and stockout avoidance, but also working-capital constraints and exposure to external trade and financial conditions. Inventory changes should consequently not be treated merely as a residual component of aggregate expenditure: they contain information about firms' capacity to absorb shocks, finance production, and coordinate activity
across periods.

\clearpage

\bibliographystyle{apalike}
\bibliography{referencias}

\clearpage

\appendix\renewcommand{\thesection}{Appendix \Alph{section}}
\section{Data-Quality Discussion for Changes in Inventories} \label{App:data_discussion}
\renewcommand{\thetable}{A\arabic{table}}
\renewcommand{\thefigure}{A\arabic{figure}}

\setcounter{table}{0}
\setcounter{figure}{0}

An important point to consider regarding the quality of the data used is that national accounts exhibit statistical discrepancies stemming from the limitations of statistical institutes in collecting and processing information, particularly in household expenditure components. However, all countries follow the guidelines of the United Nations System of National Accounts (1993 or 2008 revisions), which allows for a reasonable comparability of these statistics across countries.

In certain cases, when some variables cannot be estimated directly or completely, they are computed residually, implicitly absorbing the statistical discrepancy. An example is the case of Argentina, where in the 1990s private consumption was estimated as the difference between supply and demand, incorporating the statistical discrepancy, and subsequently, a similar criterion was applied to changes in inventories between 2004 and 2017.

The relevant issue for this paper is whether this discrepancy is large enough to affect the dynamics of the variable of interest, in this case, changes in inventories. To this end, we use World Bank statistical discrepancy data at current prices for the 1993--2022 period and construct the following ratio:

\begin{equation}
Ratio_{i,t} = \left| \frac{Disc_{i,t}}{\Delta INV_{curr, i,t}} \right|
\end{equation}

where $Disc_{i,t}$ is the statistical discrepancy and $\Delta INV_{curr, i,t}$ is the change in inventories for country $i$ in year $t$. Both variables are expressed at current prices. Based on this indicator, we calculate its distribution by country group, including the median, the proportion of observations with zero values, and the proportion of cases in which the ratio is less than 10\%.

According to the descriptive evidence, the relative incidence of the discrepancy on changes in inventories is low in most of the sample. In particular, between 57\% and almost 80\% of the observations (depending on the country group) the discrepancy is zero or represents less than 10\% of the changes in inventories. Overall, these results suggest that the statistical discrepancy operates as a residual component of limited magnitude in relation to inventory dynamics, which reduces its potential relevance as a source of distortion in the cyclical analysis of this variable at the panel level (Table \ref{tab:discrepancy_magnitude}).

\begin{table}[H]
    \centering
    \caption{Magnitude of Statistical Discrepancy Relative to Changes in Inventories (1993--2022)}
    \label{tab:discrepancy_magnitude}
    \renewcommand{\arraystretch}{1.1} 
    \begin{tabular}{l l c c c}
        \toprule
        \textbf{Classification} & \textbf{Median} & \textbf{Ratio $< 10\%$} & \textbf{Zero Disc.} \\
         &  & \textbf{(\%)} & \textbf{(\%)} \\
        \midrule
        \midrule
        Advanced & 0.0000 & 79.2 & 41.2 \\
        High-income EMs & 0.0000 & 57.1 & 27.1 \\
        Middle-income EMs & 0.0000 & 75.8 & 45.7 \\
        Low-income EMs & 0.0001 & 59.0 & 24.0 \\
        \bottomrule
    \end{tabular}
    
    \vspace{1.5ex}
    \begin{minipage}{0.95\textwidth}
        \footnotesize \textit{Notes:} Authors' calculations based on World Bank data. ``Ratio $< 10\%$ (\%)'' is the percentage of observations where the absolute discrepancy is less than 10\% of the reported changes in inventories. ``Zero Disc. (\%)'' is the percentage of observations where the statistical discrepancy is exactly zero.
    \end{minipage}
\end{table}

\newpage

\renewcommand{\thesection}{Appendix \Alph{section}}
\section{Robustness to the Inclusion of Large Economies} \label{app:large_economies}
\renewcommand{\thetable}{B\arabic{table}}
\renewcommand{\thefigure}{B\arabic{figure}}

\setcounter{table}{0}
\setcounter{figure}{0}

This appendix presents a robustness check of the main results by including and excluding large economies from the sample. In this study, large economies are defined as countries whose share of world GDP exceeds 2\%. These economies are: (i) the United States; (ii) Germany; (iii) the United Kingdom; (iv) Italy; (v) France; and (vi) Japan.
We recalculate the correlations, relative volatilities, and serial correlations of inventory dynamics with and without these economies. Overall, the results remain similar and preserve the main stylized facts discussed in the paper.

The main differences are quantitative rather than qualitative. In particular, the correlations between inventories and output, investment, exports and imports are stronger when large economies are included in the sample. By contrast, the correlations with private and public consumption become slightly weaker once these economies are incorporated.

In the case of relative volatility, the main stylized facts are also preserved. However, the magnitude of relative volatilities becomes lower for almost all variables when large economies are included in the sample, with the exception of output volatility, which remains broadly unchanged across both specifications.

Finally, the serial correlation of inventory fluctuations increases slightly when large economies are included in the sample, rising from 0.13 to 0.15. This result suggests a somewhat higher degree of persistence in inventory dynamics among large advanced economies (Table \ref{tab:advanced_large_economies}).

\begin{table}[H]
\centering
\caption{Advanced Economies: Stylized Facts with and without Large Economies}
\label{tab:advanced_large_economies}
\small
\begin{tabular}{lcc}
\toprule
\textbf{Indicator} &
\textbf{Excluding Large Economies} &
\textbf{Including Large Economies} \\
\midrule

\multicolumn{3}{l}{\textit{Correlations}} \\

Corr($\Delta inv_t^c,\Delta y_t$) & 0.48 & 0.52 \\
Corr($\Delta inv_t^c,\Delta c_t$) & 0.37 & 0.30 \\
Corr($\Delta inv_t^c,\Delta i_t$) & 0.32 & 0.34 \\
Corr($\Delta inv_t^c,\Delta g_t$) & 0.09 & 0.02 \\
Corr($\Delta inv_t^c,\Delta x_t$) & 0.33 & 0.42 \\
Corr($\Delta inv_t^c,\Delta m_t$) & 0.45 & 0.55 \\

\midrule
\multicolumn{3}{l}{\textit{Relative Volatility}} \\

$\sigma(\Delta inv_t^c)/\sigma(\Delta y_t)$ & 0.26 & 0.26 \\
$\sigma(\Delta inv_t^c)/\sigma(\Delta c_t)$ & 0.33 & 0.31 \\
$\sigma(\Delta inv_t^c)/\sigma(\Delta i_t)$ & 0.11 & 0.10 \\
$\sigma(\Delta inv_t^c)/\sigma(\Delta g_t)$ & 0.37 & 0.34 \\
$\sigma(\Delta inv_t^c)/\sigma(\Delta x_t)$ & 0.11 & 0.10 \\
$\sigma(\Delta inv_t^c)/\sigma(\Delta m_t)$ & 0.12 & 0.10 \\

\midrule
\multicolumn{3}{l}{\textit{Serial Correlation}} \\

Corr($\Delta inv_t^c,\Delta inv_{t-1}^c$) & 0.13 & 0.15 \\

\bottomrule
\end{tabular}

\vspace{0.2cm}

{\raggedright\footnotesize
Source: Authors' calculations based on World Bank data.
Large economies include the United States, Japan, Germany, France, Italy, and the United Kingdom.
\par}
\end{table}

\begin{figure}[H]
    \centering
    \caption{Long-Run Productivity-Shock Responses: Including versus Excluding Large Economies}
    \label{fig:irf_TFP_rob}
    \includegraphics[width=.9\textwidth]{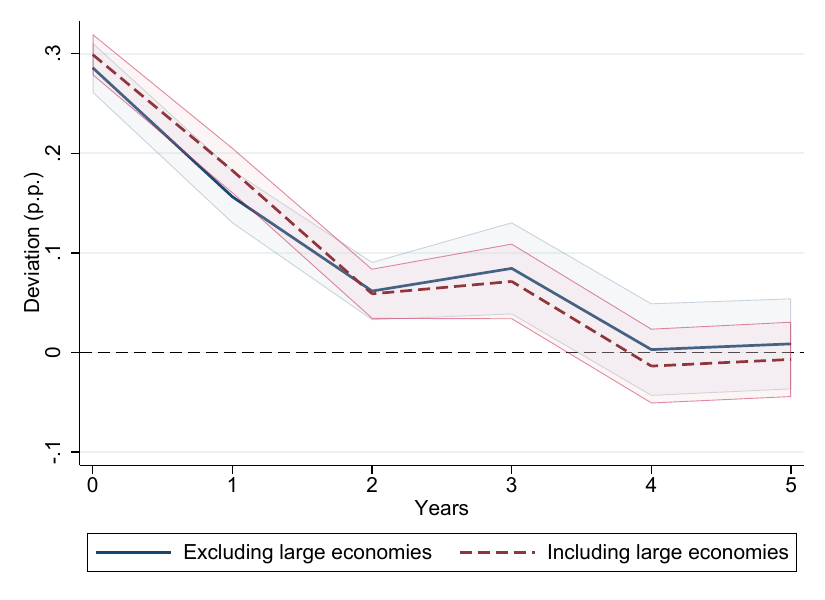}
        \begin{tabular}{p{15.0cm} p{width}}
    {\setstretch{1.1}
    \scriptsize
    \textit{Note}: Authors' elaboration based on World Bank data. Shaded areas represent 68\% confidence intervals.}
\end{tabular}
\end{figure}

Regarding the IRFs, the results obtained including and excluding large economies are broadly similar. The main difference is that, when large economies are included, the response is slightly larger on impact and in the first horizon, but becomes somewhat smaller from the second horizon onward relative to the baseline specification that excludes these economies. In addition, the confidence intervals are generally narrower, which is consistent with the larger sample size used in the estimation (Figure \ref{fig:irf_TFP_rob}).

\begin{figure}[H]
    \centering
    \caption{Financial-Shock Responses: Including versus Excluding Large Economies}
    \label{fig:irf_spread_others}
    \includegraphics[width=.9\textwidth]{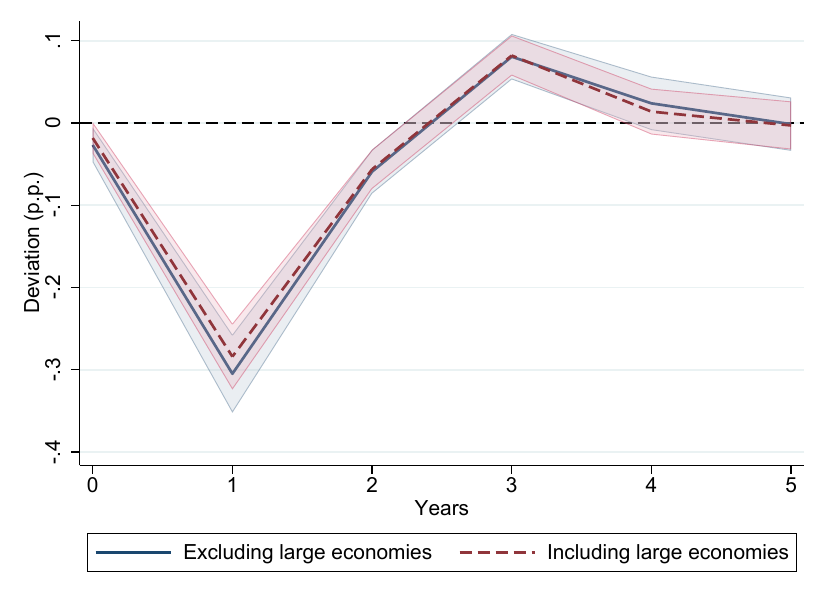}
    \begin{tabular}{p{15.0cm} p{width}}
    {\setstretch{1.1}
    \scriptsize
    \textit{Note}: Authors' elaboration based on World Bank data. Shaded areas represent 68\% confidence intervals.}
    \end{tabular}
\end{figure}

On the other hand, the results obtained including and excluding large economies in response to financial shock are also similar. The main difference is that, when large economies are included, the impact of financial shocks becomes slightly smaller from horizons zero to three compared with the baseline specification, excluding these economies. In addition, the confidence intervals become narrower, which is consistent with the larger sample size used in the estimation (Figure \ref{fig:irf_spread_others}).

\clearpage

\renewcommand{\thesection}{Appendix \Alph{section}}
\section{Robustness to Alternative Filtering Methods} \label{app:filtering_methods}
\renewcommand{\thetable}{C\arabic{table}}
\renewcommand{\thefigure}{C\arabic{figure}}

\setcounter{table}{0}
\setcounter{figure}{0}

Table \ref{tab:filters_comparison} evaluates the robustness of the main stylized facts to alternative transformations of the data. The baseline specification uses the residual from a country-specific quadratic trend for the inventory-to-trend-output ratio and log first differences for the remaining macroeconomic aggregates. We compare these results with cyclical components obtained using log-quadratic trends and the
Hodrick--Prescott (HP) filter.

Overall, the main stylized facts remain broadly robust across filtering methods. Although the magnitude of some correlations and relative volatilities changes depending on the detrending procedure, the qualitative patterns remain remarkably stable.

First, inventory fluctuations remain strongly procyclical in advanced economies under all filtering methods. Across filters, advanced economies generally display stronger correlations between inventories and output, exports, and imports. The ranking is less uniform for consumption and investment. In contrast, lower-income economies continue to display weak or near-zero correlations between inventories and macroeconomic aggregates, suggesting a lower degree of intertemporal smoothing through inventories.

Second, imports remain the macroeconomic component most strongly associated with inventory fluctuations across all filters and country groups. This result is particularly robust and supports the interpretation that inventories are closely linked to imported intermediate inputs and international trade dynamics.

Third, the volatility of inventories relative to output remains generally higher in emerging economies, especially in the lower-income groups. Regardless of the filter employed, advanced economies exhibit the lowest inventory volatility, while lower-income economies display substantially larger fluctuations relative to output and consumption. This finding is consistent with greater macroeconomic instability, financing constraints, and weaker production smoothing mechanisms in less developed economies.

Finally, the weak relationship between inventories and government consumption also remains stable across filters, indicating that inventory dynamics are more closely connected to private-sector activity and external trade conditions than to fiscal expenditure.

Taken together, these results suggest that the main conclusions of the paper do not depend on a particular detrending methodology. The broad cross-country patterns in inventory dynamics remain robust across first differences, log-quadratic detrending, and HP filtering procedures.

\begin{table}[H]
\centering
\caption{Stylized Facts under Alternative Filtering Methods}
\label{tab:filters_comparison}
\scriptsize
\renewcommand{\arraystretch}{1.1}
\setlength{\tabcolsep}{4pt}

\begin{tabular}{llccccc}
\toprule

\multicolumn{1}{c}{
    \multirow[c]{2}{*}{\textbf{Filter}}
}
&
\multicolumn{1}{c}{
    \multirow[c]{2}{*}{\textbf{Indicator}}
}
&
\multirow[c]{2}{*}{\textbf{Total}}
&
\multirow[c]{2}{*}{\textbf{Advanced}}
&
\multicolumn{3}{c}{\textbf{Emerging}} \\

\cmidrule(lr){5-7}

& & & &
\textbf{High-income} &
\textbf{Middle-income} &
\textbf{Low-income} \\

\midrule

\multicolumn{7}{l}{\textit{Panel A: Correlations}} \\

\multirow[c]{6}{*}{First Difference}
& Corr($\Delta inv_t^{c}$,$\Delta y_t$)
& 0.33 & 0.48 & 0.39 & 0.23 & -0.02 \\

& Corr($\Delta inv_t^{c}$,$\Delta c_t$)
& 0.21 & 0.37 & 0.38 & 0.17 & -0.06 \\

& Corr($\Delta inv_t^{c}$,$\Delta i_t$)
& 0.24 & 0.32 & 0.28 & 0.19 & -0.05 \\

& Corr($\Delta inv_t^{c}$,$\Delta x_t$)
& 0.16 & 0.33 & 0.23 & 0.05 & -0.02 \\

& Corr($\Delta inv_t^{c}$,$\Delta m_t$)
& 0.33 & 0.45 & 0.44 & 0.25 & 0.12 \\

& Corr($\Delta inv_t^{c}$,$\Delta g_t$)
& 0.08 & 0.09 & 0.05 & -0.04 & 0.10 \\

\midrule

\multirow[c]{6}{*}{Log-Quadratic}
& Corr($\Delta inv_t^{c}$,$y_t$)
& 0.253 & 0.487 & 0.130 & 0.158 & 0.103 \\

& Corr($\Delta inv_t^{c}$,$c_t$)
& 0.159 & 0.348 & 0.060 & 0.157 & -0.100 \\

& Corr($\Delta inv_t^{c}$,$i_t$)
& 0.231 & 0.316 & 0.127 & 0.135 & 0.163 \\

& Corr($\Delta inv_t^{c}$,$x_t$)
& 0.149 & 0.372 & 0.080 & 0.020 & 0.113 \\

& Corr($\Delta inv_t^{c}$,$m_t$)
& 0.394 & 0.518 & 0.236 & 0.314 & 0.290 \\

& Corr($\Delta inv_t^{c}$,$g_t$)
& 0.089 & 0.135 & -0.017 & 0.038 & 0.142 \\

\midrule

\multirow[c]{6}{*}{HP Filter}
& Corr($\Delta inv_t^{c}$,$y_t$)
& 0.208 & 0.518 & 0.186 & 0.173 & 0.112 \\

& Corr($\Delta inv_t^{c}$,$c_t$)
& 0.124 & 0.332 & 0.117 & 0.088 & -0.132 \\

& Corr($\Delta inv_t^{c}$,$i_t$)
& 0.154 & 0.308 & 0.048 & 0.102 & 0.044 \\

& Corr($\Delta inv_t^{c}$,$x_t$)
& 0.149 & 0.393 & 0.109 & 0.011 & 0.020 \\

& Corr($\Delta inv_t^{c}$,$m_t$)
& 0.382 & 0.501 & 0.264 & 0.357 & 0.234 \\

& Corr($\Delta inv_t^{c}$,$g_t$)
& 0.048 & 0.096 & -0.040 & -0.004 & 0.106 \\

\midrule

\multicolumn{7}{l}{\textit{Panel B: Relative Volatility}} \\

\multirow[c]{6}{*}{First Difference}
& $\sigma(\Delta inv_t^c)/\sigma(\Delta y_t)$
& 0.30 & 0.26 & 0.31 & 0.37 & 0.54 \\

& $\sigma(\Delta inv_t^c)/\sigma(\Delta c_t)$
& 0.32 & 0.33 & 0.33 & 0.32 & 0.29 \\

& $\sigma(\Delta inv_t^c)/\sigma(\Delta i_t)$
& 0.11 & 0.11 & 0.11 & 0.12 & 0.12 \\

& $\sigma(\Delta inv_t^c)/\sigma(\Delta x_t)$
& 0.14 & 0.11 & 0.15 & 0.15 & 0.10 \\

& $\sigma(\Delta inv_t^c)/\sigma(\Delta m_t)$
& 0.12 & 0.12 & 0.09 & 0.13 & 0.10 \\

& $\sigma(\Delta inv_t^c)/\sigma(\Delta g_t)$
& 0.30 & 0.37 & 0.39 & 0.27 & 0.13 \\

\midrule

\multirow[c]{6}{*}{Log-Quadratic}
& $\sigma(\Delta inv_t^c)/\sigma(y_t)$
& 26.652 & 22.654 & 25.371 & 28.494 & 40.354 \\

& $\sigma(\Delta inv_t^c)/\sigma(c_t)$
& 23.082 & 31.946 & 19.858 & 22.370 & 32.927 \\

& $\sigma(\Delta inv_t^c)/\sigma(i_t)$
& 8.021 & 8.809 & 7.979 & 7.905 & 10.886 \\

& $\sigma(\Delta inv_t^c)/\sigma(x_t)$
& 13.257 & 10.560 & 16.269 & 15.131 & 7.014 \\

& $\sigma(\Delta inv_t^c)/\sigma(m_t)$
& 10.008 & 12.005 & 8.025 & 10.459 & 8.812 \\

& $\sigma(\Delta inv_t^c)/\sigma(g_t)$
& 22.512 & 33.294 & 29.746 & 18.776 & 12.704 \\

\midrule

\multirow[c]{6}{*}{HP Filter}
& $\sigma(\Delta inv_t^c)/\sigma(y_t)$
& 37.096 & 31.977 & 37.243 & 39.198 & 61.554 \\

& $\sigma(\Delta inv_t^c)/\sigma(c_t)$
& 33.718 & 38.335 & 32.252 & 32.230 & 36.647 \\

& $\sigma(\Delta inv_t^c)/\sigma(i_t)$
& 12.235 & 12.522 & 11.946 & 11.609 & 14.275 \\

& $\sigma(\Delta inv_t^c)/\sigma(x_t)$
& 16.128 & 14.233 & 17.898 & 19.325 & 10.559 \\

& $\sigma(\Delta inv_t^c)/\sigma(m_t)$
& 12.903 & 13.044 & 10.864 & 14.274 & 11.473 \\

& $\sigma(\Delta inv_t^c)/\sigma(g_t)$
& 29.877 & 42.780 & 39.024 & 27.528 & 13.951 \\

\bottomrule
\end{tabular}

\vspace{0.2cm}

{\raggedright\footnotesize
Source: Authors' calculations based on World Bank data.
The table reports median country-level moments for inventory cycles under
three alternative detrending methods: first differences, log-quadratic
trend, and HP filter.
\par}

\end{table}

\clearpage

\renewcommand{\thesection}{Appendix \Alph{section}}
\section{Robustness of Productivity-Shock Identification to the Pooling Restriction}
\label{app:rob_productivity}

\renewcommand{\thetable}{D\arabic{table}}
\renewcommand{\thefigure}{D\arabic{figure}}
\renewcommand{\theequation}{D\arabic{equation}}

\setcounter{table}{0}
\setcounter{figure}{0}
\setcounter{equation}{0}

The baseline identification in Section~3.1 pools the adjustment coefficients $\alpha$ and the short-run dynamics $\Gamma$ across countries in order to obtain a stable estimate of the long-run impact matrix, while allowing for
country fixed effects. This appendix evaluates the sensitivity of the recovered productivity shock to that pooling restriction in three ways. First, we re-estimate the system separately for each development group. Second, we test
whether the data reject slope homogeneity within each group. Third, we implement a leave-one-group-out exercise in which the coefficients used to construct a group's comparison shock are estimated without observations from that group.

The notation follows Section~3.1. Let $\mathcal{G}_g$ denote the set of countries belonging to development group $g$, and let $\mathcal{G}_{-g}=\bigcup_{h\neq g}\mathcal{G}_h$ denote the countries belonging to the other three groups. Throughout the appendix, the $3\times2$ cointegration matrix $\beta$ remains fixed at the balanced-growth restriction imposed in Equation~(2). Only $\alpha$ and $\Gamma$ are
re-estimated. As in the main text, $\beta_{\perp}$ and $\alpha_{\perp}$ denote orthogonal complements, and $e_y'=[1\ 0\ 0]$ selects the output component.

\subsection{Group-Specific Re-estimation of the Long-Run Impact Matrix}
\label{app:group_specific_productivity}

We first re-estimate the reduced-form system separately within each development group. We keep $\beta$ fixed at the balanced-growth restriction but allow $\alpha$ and $\Gamma$ to vary across groups:

\begin{equation}
\Delta X_{i,t} = \mu_i + \alpha_g\beta'X_{i,t-1} + \sum_{j=1}^{p} \Gamma_{g,j}\Delta X_{i,t-j} + u_{i,t,g} \qquad
i\in\mathcal{G}_g
\label{eq:F1}
\end{equation}

\noindent
where $p=1$ in the baseline specification, as in the main text. The group-specific long-run impact matrix is:

\begin{equation}
\beta_{\perp}
C_g(1) = \beta_{\perp} \left[\alpha_{g,\perp}^{\prime}\left(I-sum_{j=1}^{p}\Gamma_{g,j}\right)\right]^{-1}\alpha_{g,\perp}^{\prime}
\label{eq:F2}
\end{equation}

\noindent
and the corresponding productivity shock is:

\begin{equation}
shock^{PS}_{i,t,g}=e_y'C_g(1)u_{i,t,g} \qquad i\in\mathcal{G}_g.
\label{eq:F3}
\end{equation}

Table\ref{tab:F1} reports the coefficients from the output equation and the first row of the long-run impact matrix for the panel-wide and group-specific systems. The latter contains the weights applied to the reduced-form innovations in output, consumption, and investment when constructing the productivity shock.

\begin{table}[H]
\centering
\caption{Output-Equation Coefficients and Long-Run Shock Weights:
Panel-Wide and Group-Specific Systems}
\label{tab:F1}

\scriptsize
\setlength{\tabcolsep}{3.2pt}
\renewcommand{\arraystretch}{1.12}

\begin{threeparttable}

\resizebox{\textwidth}{!}{%
\begin{tabular}{lccccccccc}
\toprule
&
&
\multicolumn{2}{c}{$\alpha_y$}
&
\multicolumn{3}{c}{$\Gamma_{y,1}$}
&
\multicolumn{3}{c}{$e_y'C_g(1)$}
\\

\cmidrule(lr){3-4}
\cmidrule(lr){5-7}
\cmidrule(lr){8-10}

\textbf{Sample}
&
\textbf{$N$}
&
\shortstack[c]{$(c-y)$}
&
\shortstack[c]{$(i-y)$}
&
\shortstack[c]{$(\Delta y)$}
&
\shortstack[c]{$(\Delta c)$}
&
\shortstack[c]{$(\Delta i)$}
&
\textbf{$C_{1,y}$}
&
\textbf{$C_{1,c}$}
&
\textbf{$C_{1,i}$}
\\

\midrule

Panel-wide (baseline)
& 72
& $-0.019$
& $-0.018$
& $-0.004$
& $0.068$
& $0.049$
& $1.444$
& $-0.249$
& $-0.128$
\\

Advanced
& 24
& $0.039$
& $-0.014$
& $0.014$
& $-0.088$
& $0.086$
& $0.329$
& \textbf{$0.861$}
& $-0.063$
\\

High-income EMs
& 14
& $0.004$
& $-0.053$
& $-0.136$
& $0.206$
& $0.035$
& $1.515$
& $-0.274$
& $-0.313$
\\

Middle-income EMs
& 24
& $-0.069$
& $-0.013$
& $0.073$
& $0.060$
& $0.036$
& $2.237$
& $-0.999$
& $-0.177$
\\

Low-income EMs
& 10
& $0.017$
& $-0.002$
& $0.066$
& $0.017$
& $0.034$
& $0.813$
& \textbf{$0.447$}
& $-0.178$
\\

\bottomrule
\end{tabular}%
}

\begin{tablenotes}
\footnotesize
\item \emph{Notes:}
Authors' calculations.
$N$ is the number of countries in the estimation sample.
$\alpha_y(c-y)$ and $\alpha_y(i-y)$ are the loadings of the output equation on the two error-correction terms $\beta'X_{i,t-1}=(c_{i,t-1}-y_{i,t-1},\,i_{i,t-1}-y_{i,t-1})'$.$\Gamma_{y,1}(\Delta y)$, $\Gamma_{y,1}(\Delta c)$, and
$\Gamma_{y,1}(\Delta i)$ are the loadings of the output equation on thelagged growth rates of output, consumption, and investment, respectively.
$C_{1,y}$, $C_{1,c}$, and $C_{1,i}$ are the elements of $e_y'C_g(1)$ and therefore the weights applied to the reduced-form innovations $u_{i,t,g}=(u_y,u_c,u_i)'$ in equation~\eqref{eq:F3}.
Boldface indicates a sign reversal relative to the corresponding panel-wide
weight.
\end{tablenotes}

\end{threeparttable}
\end{table}

The group-specific coefficients and long-run weights display meaningful variation. Emerging high-income economies most closely reproduce the panel-wide weights across the three components. The weight applied to the consumption innovation changes sign in advanced and emerging low-income economies, whereas emerging high-income and emerging middle-income economies preserve the negative sign of the panel-wide estimate. By contrast, the weight applied to the investment innovation remains negative and of a similar order of magnitude in all specifications. The construction of the shock therefore appears more sensitive to the consumption innovation than
to the investment innovation.

Despite these differences in individual coefficients and weights, the resulting shock series remain closely related. Table\ref{tab:F2} reports the correlation between the panel-wide baseline shock and the shock obtained from each group-specific system.

\begin{table}[H]
\centering
\caption{Correlation between Panel-Wide and Group-Specific Productivity Shocks}
\label{tab:F2}

\begin{threeparttable}
\begin{tabular}{lccc}
\toprule
\textbf{Group}
&
\textbf{Countries}
&
\textbf{Observations}
&
\shortstack[c]{\textbf{Correlation with}\\
\textbf{group-specific shock}}
\\
\midrule

Advanced
& 24 & 672 & $0.800$ \\

High-income EMs
& 14 & 392 & $0.897$ \\

Middle-income EMs
& 24 & 672 & $0.918$ \\

Low-income EMs
& 10 & 280 & $0.789$ \\

\bottomrule
\end{tabular}

\begin{tablenotes}[flushleft]
\footnotesize
\item \emph{Notes:}
The table reports
$\operatorname{Corr}
(\text{shock}^{PS}_{i,t},\text{shock}^{PS}_{i,t,g})$,
where $\text{shock}^{PS}_{i,t}$ is obtained from the panel-wide system in equation~(5) and $\text{shock}^{PS}_{i,t,g}$ is obtained from the group-specific system in Equation\eqref{eq:F3}. Correlations are calculated
over the country-year observations belonging to each group. The correlation with the sign-normalized structural shock is numerically identical up to a positive rescaling and is therefore not reported separately.
\end{tablenotes}

\end{threeparttable}
\end{table}

All four correlations are close to or above $0.80$, with the lowest value equal to $0.789$ for emerging low-income economies. This indicates that allowing the adjustment coefficients and short-run dynamics to vary by development group changes some elements of the long-run mapping but does not produce an entirely different productivity-shock series.

\begin{figure}[H]
    \centering
    \caption{Inventory Responses to the Productivity Shock:
    Panel-Wide and Group-Specific Reconstructions}
    \label{fig:F1}

    \includegraphics[width=\textwidth]
    {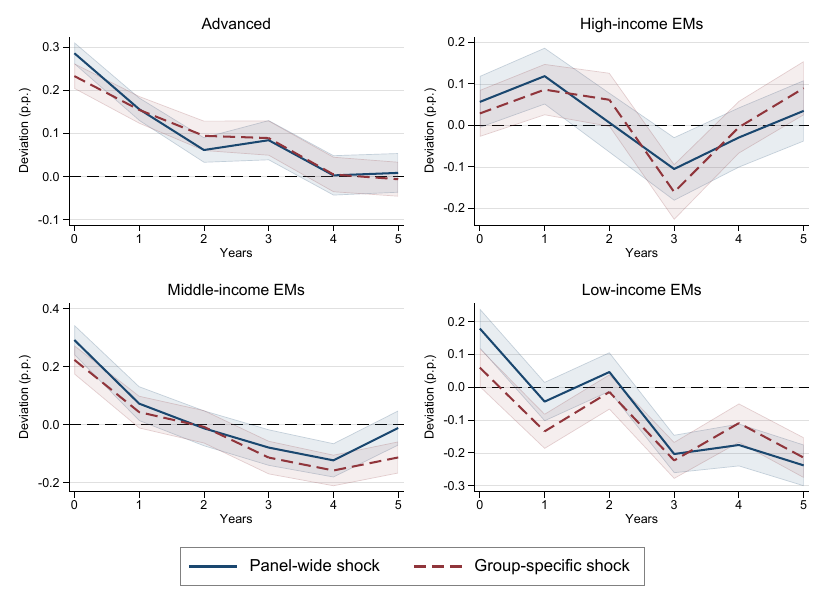}

    \vspace{0.15cm}

    \begin{minipage}{0.95\textwidth}
        \setstretch{1.1}
        \scriptsize
        \textit{Notes:}
        The figure compares the inventory responses obtained using the
        panel-wide productivity shock and the corresponding group-specific
        reconstruction. Shaded areas represent 68\% confidence intervals.
        Source: Authors' calculations based on World Bank data.
    \end{minipage}
\end{figure}

Figure\ref{fig:F1} compares the inventory responses obtained with the group-specific shocks with the baseline responses reported in Figure~2. We use the same local-projection specification in both cases.

The two specifications follow similar paths in advanced and emerging high-income economies. They also preserve the main sign and adjustment patterns in emerging middle-income and emerging low-income economies, although some differences arise in timing and magnitude. The correlations in Table\ref{tab:F2}, however, compare each group-specific shock with a panel-wide system that includes observations from the same group. The
leave-one-group-out exercise in \ref{app:leave_group_out} provides a stricter assessment by excluding each group from the estimation of the
comparison system.

\subsection{Slope-Homogeneity Tests}
\label{app:slope_homogeneity}

We next evaluate whether the data reject the restriction of common short-run dynamics. We use the slope-homogeneity test of \citet{pesaran2008testing}, applied separately to the output, consumption, and investment equations of the reduced-form system.

For each equation, the two error-correction coefficients are maintained as common within the sample being tested, in line with the baseline specification. The homogeneity test concerns the three coefficients on the lagged growth rates collected in $\Gamma_1$. Accordingly, let $\hat b_i$ denote the country-specific estimate of these $k=3$ short-run coefficients
and let $\hat b_{\mathrm{pool}}$ denote the corresponding pooled estimate.
The test statistic is based on:

\begin{equation}
S=\sum_{i=1}^{N}\left(\hat b_i-\hat b_{\mathrm{pool}}\right)^{\prime}
\widehat{\operatorname{Var}}(\hat b_i)^{-1}
\left(\hat b_i-\hat b_{\mathrm{pool}}\right)
\label{eq:F4}
\end{equation}

\noindent
from which the standard and small-sample-adjusted statistics are calculated as:

\begin{equation}
\Delta=\sqrt{N}\frac{(S/N)-k}{\sqrt{2k}},
\qquad
\Delta_{\mathrm{adj}}
=\sqrt{N}\frac{(S/N)-k}{\sqrt{2k(\bar T-k-1)/(\bar T+1)}}
\label{eq:F5}
\end{equation}

\noindent
where $\bar T$ is the average number of time-series observations per country.
Under the null hypothesis of slope homogeneity, both statistics are asymptotically standard normal. We base the summary in Table\ref{tab:F3} on $\Delta_{\mathrm{adj}}$. Allowing the error-correction coefficients to vary across countries yields similar qualitative conclusions and is not reported here.

\begin{table}[H]
\centering
\caption{Slope-Homogeneity Tests for the Short-Run Coefficients}
\label{tab:F3}
\begin{threeparttable}
\begin{tabular}{lccccc}
\toprule
\textbf{Sample} & \textbf{$N$} & \textbf{$\Delta$} & \textbf{$p(\Delta)$} & \textbf{$\Delta_{\mathrm{adj}}$} & \textbf{$p(\Delta_{\mathrm{adj}})$} \\
\midrule
\multicolumn{6}{l}{\textit{Panel A: Output equation ($\Delta y$)}} \\
\midrule
Full sample             & 72 & $3.7895$  & $0.0002$ & $4.1656^{***}$ & $0.0000$ \\
Advanced                & 24 & $0.3628$  & $0.7167$ & $0.3988$        & $0.6900$ \\
High-income EMs    & 14 & $0.1434$  & $0.8859$ & $0.1577$        & $0.8747$ \\
Middle-income EMs  & 24 & $3.4635$  & $0.0005$ & $3.8072^{***}$ & $0.0001$ \\
Low-income EMs    & 10 & $2.8731$  & $0.0041$ & $3.1583^{***}$ & $0.0016$ \\
\midrule
\multicolumn{6}{l}{\textit{Panel B: Consumption equation ($\Delta c$)}} \\
\midrule
Full sample             & 72 & $7.7759$  & $0.0000$ & $8.5477^{***}$ & $0.0000$ \\
Advanced                & 24 & $6.6185$  & $0.0000$ & $7.2753^{***}$ & $0.0000$ \\
High-income EMs    & 14 & $0.3538$  & $0.7235$ & $0.3890$        & $0.6973$ \\
Middle-income EMs  & 24 & $5.3336$  & $0.0000$ & $5.8630^{***}$ & $0.0000$ \\
Low-income EMs    & 10 & $1.0014$  & $0.3166$ & $1.1008$        & $0.2710$ \\
\midrule
\multicolumn{6}{l}{\textit{Panel C: Investment equation ($\Delta i$)}} \\
\midrule
Full sample             & 72 & $8.2102$  & $0.0000$ & $9.0250^{***}$ & $0.0000$ \\
Advanced                & 24 & $5.2801$  & $0.0000$ & $5.8041^{***}$ & $0.0000$ \\
High-income EMs    & 14 & $-0.2610$ & $0.7941$ & $-0.2870$       & $0.7741$ \\
Middle-income EMs  & 24 & $5.5664$  & $0.0000$ & $6.1189^{***}$ & $0.0000$ \\
Low-income EMs    & 10 & $1.6529$  & $0.0983$ & $1.8170^{*}$    & $0.0692$ \\
\bottomrule
\end{tabular}
\begin{tablenotes}[flushleft]
\footnotesize
\item \emph{Notes:} $N$ is the number of countries in the sample. $\Delta$ and $\Delta_{\mathrm{adj}}$ are the Pesaran--Yamagata slope-homogeneity statistics defined in equations~\eqref{eq:F4}--\eqref{eq:F5}, testing the null that the $k=3$ short-run coefficients in $\Gamma_1$ (the loadings on $\Delta y_{t-1}$, $\Delta c_{t-1}$, $\Delta i_{t-1}$) are common across countries in the sample; $p(\Delta)$ and $p(\Delta_{\mathrm{adj}})$ are the corresponding two-sided $p$-values under the asymptotic standard normal null. $\bar T = 28$ years in every sample, reflecting the balanced panel. The error-correction terms are pooled together with $\Gamma_1$ in every specification, mirroring the restriction imposed in equation~\eqref{eq:F1}; results are nearly identical under a less restrictive specification in which the error-correction coefficients are allowed to vary freely by country, reported in the replication package together with the underlying test program. $^{*}$, $^{**}$, $^{***}$ denote rejection of the null of slope homogeneity, based on $\Delta_{\mathrm{adj}}$, at the 10\%, 5\%, and 1\% level, respectively.
\end{tablenotes}
\end{threeparttable}
\end{table}

The null of slope homogeneity is rejected in all three equations for the full sample, indicating that short-run dynamics are not homogeneous across the 72 countries taken together. Within-group results are more varied.
Homogeneity is not rejected in any equation for emerging high-income economies, which is also the group whose long-run weights most closely resemble the panel-wide estimates.

Emerging middle-income economies reject homogeneity in all three equations, the strongest rejection pattern among the four groups. Advanced economies reject homogeneity in the consumption and investment equations but not in
the output equation. Because $C_g(1)$ depends on the complete reduced-form system, homogeneity in the output equation alone does not guarantee that the reconstructed shock will be insensitive to the remaining equations.
Emerging low-income economies also reject homogeneity in two equations, but these results should be interpreted cautiously given the limited power of the test with only ten countries.

\subsection{Leave-One-Group-Out Validation}
\label{app:leave_group_out}

The comparison in Table\ref{tab:F2} uses a panel-wide baseline that contains observations from the group against which it is compared. This may matter particularly for advanced and emerging middle-income economies, which are the two largest groups in the sample. We therefore implement a stricter out-of-group validation.

For each group $g$, we re-estimate the reduced-form system using only the countries belonging to $\mathcal{G}_{-g}$:

\begin{equation}
\Delta X_{i,t}=\mu_i+\alpha_{-g}\beta'X_{i,t-1}+\sum_{j=1}^{p}\Gamma_{-g,j}\Delta X_{i,t-j}+u_{i,t,-g},
\qquad i\in\mathcal{G}_{-g}.
\label{eq:F6}
\end{equation}

The resulting long-run impact matrix is

\begin{equation}
C_{-g}(1)=\beta_{\perp}\left[\alpha_{-g,\perp}^{\prime}\left(
I-\sum_{j=1}^{p}\Gamma_{-g,j}\right)\beta_{\perp}\right]^{-1}
\alpha_{-g,\perp}^{\prime}
\label{eq:F7}
\end{equation}

Thus, observations from group $g$ do not enter the estimation of $\alpha_{-g}$, $\Gamma_{-g}$, or $C_{-g}(1)$. We then apply these externally estimated slope coefficients to each country in the excluded group. Its fixed effect is estimated from its own time series as:

\begin{equation}
\hat\mu_i=\frac{1}{T_i}\sum_{t=1}^{T_i}\left[\Delta X_{i,t}
-\alpha_{-g}\beta^{\prime}X_{i,t-1}-\sum_{j=1}^{p}\Gamma_{-g,j}\Delta X_{i,t-j}\right] \qquad i\in\mathcal{G}_g.
\label{eq:F8}
\end{equation}

The leave-one-group-out productivity shock is then:

\begin{equation}
\begin{split}
shock^{LOO}_{i,t,g}=e_y^{\prime}C_{-g}(1)\Bigg[
&\Delta X_{i,t}-\hat\mu_i-\alpha_{-g}\beta^{\prime}X_{i,t-1}\\
&-\sum_{j=1}^{p}\Gamma_{-g,j}\Delta X_{i,t-j}\Bigg] \qquad
i\in\mathcal{G}_g
\end{split}
\label{eq:F9}
\end{equation}

The long-run mapping applied to group $g$ is therefore estimated entirely from the other three groups. Observations from group $g$ are used only to estimate its country fixed effects, recover its reduced-form innovations, and construct the corresponding comparison shock.

Table~\ref{tab:F4} reports the size and long-run weights of each leave-one-group-out system.

\begin{table}[H]
\centering
\caption{Leave-One-Group-Out System Diagnostics}
\label{tab:F4}

\begin{threeparttable}
\begin{tabular}{lcccc}
\toprule
\textbf{Group excluded}
&
\shortstack[c]{\textbf{Countries in}\\$\boldsymbol{\mathcal{G}_{-g}}$}
&
\textbf{$C_{-g,1,y}$}
&
\textbf{$C_{-g,1,c}$}
&
\textbf{$C_{-g,1,i}$}
\\
\midrule

Advanced
& 48 & $1.956$ & $-0.715$ & $-0.175$ \\

High-income EMs
& 58 & $1.401$ & $-0.217$ & $-0.089$ \\

Middle-income EMs
& 48 & $0.817$ & \textbf{$0.388$} & $-0.146$ \\

Emerging low-income EMs
& 62 & $1.489$ & $-0.298$ & $-0.140$ \\

\bottomrule
\end{tabular}

\begin{tablenotes}[flushleft]
\footnotesize
\item \emph{Notes:}
$C_{-g,1,y}$, $C_{-g,1,c}$, and $C_{-g,1,i}$ are the elements of
$e_y^{\prime}C_{-g}(1)$ obtained from the countries in $\mathcal{G}_{-g}$ using Equations\eqref{eq:F6}--\eqref{eq:F7}. Boldface indicates a sign reversal relative to the corresponding group-specific weight reported in
Table~\ref{tab:F1}.
\end{tablenotes}

\end{threeparttable}
\end{table}

Table~\ref{tab:F5} compares the baseline correlations in
Table~\ref{tab:F2} with the correlations obtained under the
leave-one-group-out design.

\begin{table}[H]
\centering
\caption{Baseline and Leave-One-Group-Out Correlations with the
Group-Specific Shock}
\label{tab:F5}

\begin{threeparttable}
\begin{tabular}{lcccc}
\toprule
\textbf{Group}
&
\textbf{Observations}
&
\textbf{Baseline}
&
\shortstack[c]{\textbf{Leave-one-}\\\textbf{group-out}}
&
\textbf{Difference}
\\
\midrule

Advanced
& 672 & $0.800$ & $0.734$ & $-0.066$ \\

High-income EMs
& 392 & $0.897$ & $0.856$ & $-0.041$ \\

Middle-income EMs
& 672 & $0.918$ & \textbf{$0.597$} & \textbf{$-0.321$} \\

Emerging low-income EMs
& 280 & $0.789$ & $0.765$ & $-0.024$ \\

\bottomrule
\end{tabular}

\begin{tablenotes}[flushleft]
\footnotesize
\item \emph{Notes:}
``Baseline'' reports
$\operatorname{Corr}
(\text{shock}^{PS}_{i,t},\text{shock}^{PS}_{i,t,g})$
from Table~\ref{tab:F2}. ``Leave-one-group-out'' reports
$\operatorname{Corr}
(\text{shock}^{PS}_{i,t,g},\text{shock}^{LOO}_{i,t,g})$
using equation~\eqref{eq:F9}. The final column reports the difference between the two correlations. Boldface identifies the case in which the decline exceeds $0.10$.
\end{tablenotes}

\end{threeparttable}
\end{table}

Three of the four groups retain correlations above $0.73$ under the leave-one-group-out exercise, with declines of between $0.024$ and $0.066$ relative to the baseline comparison. The reconstructed shocks for advanced, emerging high-income, and emerging low-income economies therefore remain closely related to their group-specific measures even when their observations
are excluded from the estimation of the comparison system.

Emerging middle-income economies are the main exception. Their correlation falls from $0.918$ to $0.597$, a decline substantially larger than that observed for the remaining groups. This result is not explained by the size
of the leave-one-group-out estimation sample alone. Excluding advanced economies also leaves a system of 48 countries, but the corresponding correlation declines by only $0.066$.

The larger decline for emerging middle-income economies is consistent with the slope-homogeneity results in Table~\ref{tab:F3}. This is the only group that rejects homogeneity in all three equations of the reduced-form system.
Its short-run dynamics therefore differ more broadly from those estimated using the remaining countries. Excluding the group changes not only the number of observations but also the dynamics used to reconstruct its comparison shock.

\subsection{Summary}
\label{app:pooling_summary}

Taken together, the results indicate that the baseline productivity shock is generally robust to the pooling of $\alpha$ and $\Gamma$, although the degree of robustness varies across groups. For advanced, emerging high-income, and
emerging low-income economies, the group-specific and leave-one-group-out shocks remain strongly correlated with the baseline measure, and the main impulse-response patterns are preserved. Emerging high-income economies provide the clearest support for the pooling approximation: slope homogeneity is not rejected in any equation, and both correlation exercises yield values
above $0.85$.

Emerging middle-income economies are the main exception. This group rejects slope homogeneity in all three equations, and the correlation with its group-specific shock falls from $0.918$ when the group is included in the comparison system to $0.597$ when it is excluded. These findings do not invalidate the productivity-shock estimates for this group, but they show
that its recovered shock is more sensitive to the pooling restriction than those of the other groups. We therefore interpret the exact magnitude and medium-horizon dynamics of the corresponding results in Section~4.1 with additional caution.

\clearpage

\section{Long-History Single-Country Validation}
\addcontentsline{toc}{section}
{Appendix E Long-History Single-Country Validation}
\label{app:long_history_validation}

\renewcommand{\thetable}{E\arabic{table}}
\renewcommand{\thefigure}{E\arabic{figure}}
\renewcommand{\theequation}{E\arabic{equation}}

\setcounter{table}{0}
\setcounter{figure}{0}
\setcounter{equation}{0}

\ref{app:rob_productivity} evaluates the sensitivity of the
productivity shock to pooling the adjustment coefficients $\alpha$ and the short-run dynamics $\Gamma$ within the panel's 1993--2022 window. This appendix provides a complementary validation using long national histories and no cross-country pooling. The exercise asks whether country-specific systems that produce well-behaved and comparable long-run mappings recover productivity shocks similar to the panel-wide measure.

Unlike the exercises in \ref{app:rob_productivity}, the
country-specific systems use several decades of pre-1993 information that do not enter the panel estimation. The exercise therefore provides a validation that is independent of cross-country pooling and relies on substantial information from outside the panel period.

\subsection*{Appendix E.1 \quad Data and Country Selection and estimation}

We extend the World Bank national accounts series for GDP, government consumption, private consumption, gross fixed capital formation, and population back to 1960 whenever continuous data are available. These are the same indicators used to construct the variables in the main panel.

Within each development group, we select countries that have continuous availability of all five series. This procedure
provides 19 countries with between 41 and 65 usable annual observations after differencing and lagging. The first available years range from 1960 to 1982, while the final years range from 2023 to 2024.

For each selected country $i$, we estimate the reduced-form system using only that country's own data. As in the main text, we keep $\beta$ fixed at the balanced-growth restriction in equation~(2), but we impose no cross-country pooling on $\alpha$ or $\Gamma$:

\begin{equation}
\Delta X_{i,t}=\mu_i+\alpha_i\beta^{\prime}X_{i,t-1}+\Gamma_{i,1}\Delta X_{i,t-1}+u_{i,t}
\label{eq:E1}
\end{equation}

\noindent
where the lag order is $p=1$, as in the baseline panel specification.
Because each system contains only one country, $\mu_i$ enters as an ordinary intercept rather than through a set of country fixed effects. It represents the same country-specific level component as $\mu_i$ in equation~(3).

The country-specific long-run impact matrix is:

\begin{equation}
C_i(1)=\beta_\perp\left[\alpha_{i,\perp}^{\prime}\left(I-\Gamma_{i,1}\right)\beta_\perp\right]^{-1}\alpha_{i,\perp}^{\prime}
\label{eq:E2}
\end{equation}

\noindent
and the corresponding productivity shock:

\begin{equation}
shock^{PS,\mathrm{country}}_{i,t}=e_y^{\prime}C_i(1)u_{i,t},
\qquad
e_y^{\prime}=[1\ 0\ 0].
\label{eq:E3}
\end{equation}

We construct the country-specific shock over the country's full available history. We then calculate its correlation with the panel-wide productivity shock in Equation (5) over the years in which both measures are available.
For nearly all countries, the common comparison period is 1995--2022, after accounting for the observations lost through differencing and lagging.

\subsection*{Appendix E.3 \quad Comparability of the Country-Specific Reconstructions}

The purpose of this exercise is not to determine whether every unrestricted single-country VECM can recover a sufficiently stable long-run shock.
Estimating $\alpha_i$, $\Gamma_i$, and $C_i(1)$ from one annual national time series is considerably more demanding than estimating the regularized panel-wide system. Rather, the exercise asks whether, when a long national history produces a well-behaved long-run mapping comparable to the one used
in the panel, the resulting country-specific shock resembles the panel-wide measure.

The country-specific long-run impact matrix depends inversely on the scalar:

\begin{equation}
M_i=\alpha_{i,\perp}^{\prime}\left(I-\Gamma_{i,1}\right)\beta_\perp.
\label{eq:G4}
\end{equation}

When $M_i$ is close to zero, small changes in the estimated country-specific coefficients can produce very large changes in $C_i(1)$. Some unrestricted country-specific systems also produce an output-innovation weight whose orientation differs from that of the panel-wide reconstruction. Such systems
do not provide directly comparable national benchmarks for this validation exercise.

\subsection*{Appendix E.4 \quad Results}

Among the 19 comparable country-specific systems, the median correlation between the long-history national shock and the panel-wide shock is $0.829$, while the mean correlation is $0.766$. Seventeen of the nineteen correlations exceed $0.5$, and only two are below $0.3$.

\begin{table}[H]
\centering
\caption{Long-History Validation Using Comparable Single-Country
Reconstructions}
\label{tab:G2}

\scriptsize
\setlength{\tabcolsep}{4pt}
\renewcommand{\arraystretch}{1.08}

\begin{threeparttable}

\resizebox{\textwidth}{!}{%
\begin{tabular}{llcccc}
\toprule
\textbf{Country}
&
\textbf{Group}
&
\textbf{Coverage}
&
\shortstack[c]{\textbf{Usable}\\\textbf{observations}}
&
\textbf{$C_{i,1,y}$}
&
\shortstack[c]{\textbf{Correlation with}\\\textbf{panel-wide shock}}
\\
\midrule

Hong Kong SAR
& Advanced
& 1961--2024
& 62
& $1.634$
& $0.949$
\\

Greece
& Advanced
& 1960--2023
& 62
& $2.435$
& $0.879$
\\

Netherlands
& Advanced
& 1969--2023
& 53
& $0.270$
& $0.873$
\\

Switzerland
& Advanced
& 1970--2023
& 52
& $0.656$
& $0.796$
\\

Australia
& Advanced
& 1960--2024
& 63
& $1.526$
& $0.233$
\\

Mexico
& High-income EMs
& 1960--2024
& 63
& $1.268$
& $0.954$
\\

Malaysia
& High-income EMs
& 1960--2024
& 63
& $1.035$
& $0.895$
\\

South Africa
& High-income EMs
& 1960--2024
& 63
& $1.116$
& $0.816$
\\

Chile
& High-income EMs
& 1960--2024
& 63
& $2.654$
& $0.026$
\\

Argentina
& High-income EMs
& 1935--2021
& 85
& $1.885$
& $0.592$
\\

Indonesia
& Middle-income EMs
& 1960--2024
& 63
& $1.670$
& $0.981$
\\

Ecuador
& Middle-income EMs
& 1960--2024
& 63
& $1.089$
& $0.965$
\\

Bolivia
& Middle-income EMs
& 1960--2024
& 63
& $2.440$
& $0.954$
\\

Dominican Republic
& Middle-income EMs
& 1960--2024
& 63
& $0.538$
& $0.891$
\\

Peru
& Middle-income EMs
& 1960--2024
& 63
& $3.632$
& $0.829$
\\

Colombia
& Middle-income EMs
& 1965--2024
& 58
& $2.042$
& $0.824$
\\

Honduras
& Middle-income EMs
& 1960--2024
& 63
& $3.321$
& $0.684$
\\

Mali
& Low-income EMs
& 1980--2024
& 43
& $1.110$
& $0.786$
\\

Uganda
& Low-income EMs
& 1982--2024
& 41
& $2.742$
& $0.626$
\\

\midrule

\multicolumn{5}{l}{\textit{Median}}
& $0.829$
\\

\multicolumn{5}{l}{\textit{Mean}}
& $0.766$
\\

\bottomrule
\end{tabular}%
}

\begin{tablenotes}[flushleft]
\footnotesize
\item \emph{Notes:}
The table reports the 19 country-specific systems satisfying the comparability criterion $0<C_{i,1,y}<5$.
Correlations are calculated between the panel-wide productivity shock in equation~(5) and the country-specific shock in equation~\eqref{eq:E3} over the years in which both measures are available.
\end{tablenotes}

\end{threeparttable}
\end{table}

The results show that, when the national history supports a comparable country-specific long-run mapping, the recovered productivity shock is generally closely related to the panel-wide measure. The relationship is particularly strong among emerging middle-income economies where six of the seven
correlations exceed $0.8$, and four exceed $0.9$. High correlations also appear in both advanced and emerging high-income economies.

Two comparable country-specific systems produce correlations below $0.3$. Chile's national reconstruction remains weakly correlated with the panel-wide shock despite producing a well-behaved long-run mapping. This suggests that its historical country-specific dynamics are not closely represented by the pooled shock. Australia's result is more sensitive to the
estimation window: although the full-history system satisfies the comparability criterion, subperiod estimates reported in the replication package indicate greater instability during the panel period. We retain both countries in the validation sample, confirming that comparability is determined independently of the correlation outcome.

Taken together with \ref{app:rob_productivity}, the long-history exercise provides two complementary forms of evidence on the pooling restriction used to construct the productivity shock.

\ref{app:rob_productivity} evaluates the shock within the panel's 1993--2022 window. Its group-specific reconstructions remain strongly correlated with the baseline measure in all four groups. Under the stricter leave-one-group-out exercise, the correlations remain above $0.73$ for advanced, emerging high-income, and emerging low-income economies, while
emerging middle-income economies display greater sensitivity to the pooling restriction.

The present exercise evaluates the identification without cross-country pooling and using each country's own extended history. Among the 19 countries for which the unrestricted single-country VECM produces a long-run mapping comparable in orientation and scale to the panel-wide system, the median
correlation with the panel shock is $0.829$, the mean is $0.766$, and 17 correlations exceed $0.5$. These results use substantial information from before the panel period and are obtained without imposing common country-level adjustment or short-run coefficients.

Overall, the evidence indicates that, when sufficiently long national series support a well-behaved and comparable country-specific reconstruction, the resulting productivity shock is generally similar to the measure recovered from the pooled panel. This finding supports the use of pooling as a
regularization device in the shorter 1993--2022 panel, while acknowledging that the panel-wide measure cannot reproduce every country's historical dynamics equally closely.

\clearpage

\renewcommand{\thesection}{Appendix \Alph{section}}
\section{Robustness to Alternative Local-Projection Specifications and Shock Diagnostics} \label{app:specifications}
\renewcommand{\thetable}{F\arabic{table}}
\renewcommand{\thefigure}{F\arabic{figure}}

\setcounter{table}{0}
\setcounter{figure}{0}

\subsection{Weak Instrument Diagnostics for the Financial Shock}
\label{app:weak_iv_financial}

We compute the statistic separately for each country group and projection horizon in the LP-IV specifications. Since the financial shock and the excluded instruments vary only over time, we cluster standard errors by year and partial out country fixed effects. The excluded instruments are Federal Reserve monetary surprises and changes in global macro-financial uncertainty (Table \ref{tab:mop_effective_f}).

\begin{table}[H]
\centering
\caption{Montiel Olea--Pflueger Effective F-Statistics}
\label{tab:mop_effective_f}
\small
\begin{tabular}{lcccccc}
\toprule
\textbf{Group} & \textbf{$h=0$} & \textbf{$h=1$} & \textbf{$h=2$} & \textbf{$h=3$} & \textbf{$h=4$} & \textbf{$h=5$} \\
\midrule
Advanced        & 10.01 & 9.75 & 11.31 & 10.14 & 12.25 & 14.85 \\
High-income EMs       & 10.31 & 10.23 & 11.05 & 9.90  & 11.84 & 14.42 \\
Middle-income EMs       & 9.54  & 9.47  & 10.77 & 9.71  & 11.50 & 14.30 \\
Low-income EMs         & 8.83  & 9.00  & 10.13 & 9.13  & 10.77 & 13.03 \\
\bottomrule
\end{tabular}
\begin{flushleft}
\footnotesize \textit{Notes:} Authors' elaboration based on World Bank data. The statistics generally exceed the critical values associated with a 20 percent maximal relative bias of TSLS, although they do not uniformly exceed the more stringent 10 percent threshold.
\end{flushleft}
\end{table}

Overall, the diagnostics indicate that the excluded instruments display moderate relevance. The effective $F$-statistics are not uniformly high enough to rule out weak-instrument concerns under the most stringent thresholds, but they generally exceed the critical values associated with a 20 percent maximal relative bias of TSLS.

\subsection{Robustness to Crisis Dummy Variables}
\label{app:specifications_2}
To ensure that the financial shock does not capture the effects of other global disturbances, we perform a robustness check using crisis dummy variables. Because of perfect collinearity with the financial shock, we cannot
include year fixed effects. Instead, we include indicators for the Asian--Russian crisis (1998), the global financial crisis (2009), and the COVID--19 pandemic (2020) in the estimation\footnote{Although the Asian-Russian crisis originated in 1997 and the global financial crisis began in 2008, their main effects on economic activity materialized in 1998 and 2009, respectively. Therefore, the corresponding crisis dummies take the value one in those years and zero otherwise.}. We estimate the local projections separately for each country group according to the following specification:

\begin{equation}\tag{D.1}
    \label{eq:dummies}
    \Delta inv^{c}_{i,t+h} = \alpha_i + \beta^{F}_h \widehat{BAA}_{t} + \Gamma^{F}_h Z_{i,t-1} + \delta_{1,h} Cri98_t + \delta_{2,h} Cri09_t + \delta_{3,h} Cri20_t +  \varepsilon^{F}_{i,t+h}
\end{equation}
where $\Delta inv_{i,t+h}^c$ represents the cyclical component of inventory changes at horizon $t+h$, and $\widehat{BAA}_{t}$ denotes the instrumented financial shock. The term $\alpha_i$ represents country fixed effects. The coefficient $\beta^{F}_h$ measures the dynamic response of inventories to the financial shock. The term $Z_{i,t-1}$ is a vector of lagged control variables containing the first and second lags of GDP growth and the inventory cycle, and $\Gamma^{F}_h$ is the vector of coefficients associated with these controls. In addition, $Cri98_t$, $Cri09_t$ and $Cri20_t$ are dummy variables that take a value of one in 1998, 2009 and 2020, respectively, and zero otherwise. Their corresponding coefficients are $\delta_{1,h}$, $\delta_{2,h}$ and $\delta_{3,h}$. Finally, $\varepsilon^{F}_{i,t+h}$ denotes the error term.

Overall, the results remain largely unchanged. Across all country groups, the sign, timing, and persistence of the inventory response to financial shocks are similar to those obtained in the baseline specification.

For advanced economies, the inventory contraction following a financial shock remains concentrated in the first two years, followed by a positive adjustment phase. Similarly, upper-middle- and high-income emerging economies, as well as lower-middle-income emerging economies, continue to exhibit a negative response at horizon one and a subsequent rebound in later periods. Finally, low-income emerging economies remain the only group displaying a persistent negative response since horizon three. In addition, the statistical significance of the main responses is preserved across specifications.

\begin{figure}[H]
    \centering
    \caption{Robustness Check Inventory Response to Financial Shock}
    \label{fig:irf_spread_rb}
    \includegraphics[width=.9\textwidth]{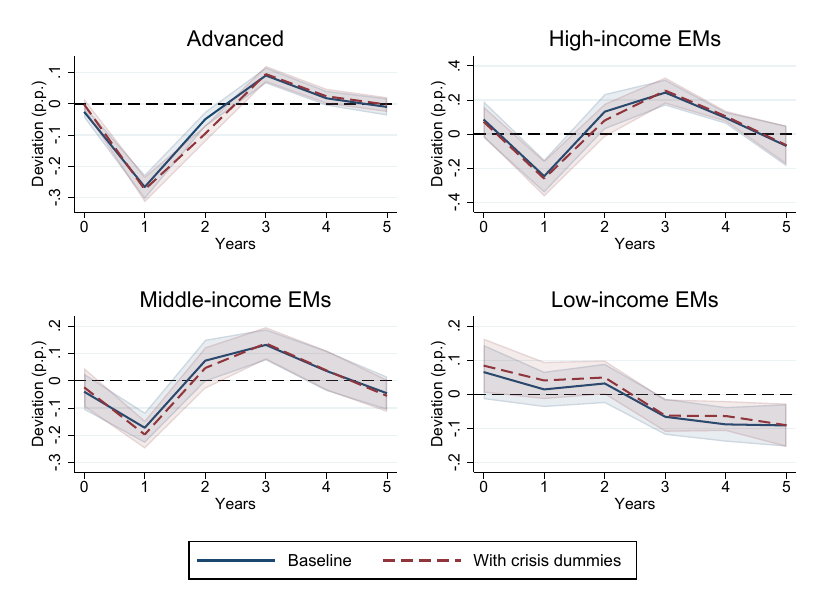}
        \begin{tabular}
    {p{15.0cm} p{width}}
    {\setstretch{1.1}
    \scriptsize
    \textit{Note}: Authors' elaboration based on World Bank data. Shaded areas represent 68\% confidence intervals.}
\end{tabular}
\end{figure}

Although some coefficients differ in magnitude, particularly at specific horizons, the inclusion of crisis dummies does not alter the qualitative interpretation of the results. This suggests that the estimated inventory responses are not driven by a small number of major global crisis episodes (Figure \ref{fig:irf_spread_rb}).

\subsection{Robustness of the Financial Shock: Global Inference and Crisis Controls}

Since the financial shock is global and varies only over time, we assess the robustness of the inference to alternative clustering schemes. In addition to the baseline country-clustered specification, we report results with year-clustered standard errors, two-way clustering by country and year, and two-way clustering combined with global crisis dummies. The main dynamic pattern remains stable across specifications; that is, the financial shock generates an inventory decumulation in the first horizon for advanced, high and upper-middle-income emerging, and lower-middle-income emerging economies, followed by a partial rebuilding of inventories at medium horizons (Table \ref{tab:financial_shock_robustness}).

\begin{table}[H]
\centering
\caption{Robustness of Financial-Shock Responses}
\label{tab:financial_shock_robustness}
\small
\begin{tabular}{llrrrrrr}
\toprule
\textbf{Group} & \textbf{Specification} & \textbf{$h=0$} & \textbf{$h=1$} & \textbf{$h=2$} & \textbf{$h=3$} & \textbf{$h=4$} & \textbf{$h=5$} \\
\midrule
Advanced & Baseline & -0.027 & -0.305 & -0.059 & 0.081 & 0.024 & -0.001 \\
Advanced & Two-way + crisis & -0.004 & -0.308 & -0.105 & 0.089 & 0.031 & 0.008 \\
High-income EMs & Baseline & 0.089 & -0.277 & 0.087 & 0.210 & 0.080 & -0.090 \\
High-income EMs & Two-way + crisis & 0.072 & -0.288 & 0.040 & 0.222 & 0.089 & -0.088 \\
Middle-income EMs & Baseline & -0.029 & -0.159 & 0.077 & 0.129 & 0.032 & -0.048 \\
Middle-income EMs & Two-way + crisis & -0.015 & -0.186 & 0.049 & 0.138 & 0.037 & -0.058 \\
Low-income EMs & Baseline & 0.065 & 0.017 & 0.024 & -0.084 & -0.103 & -0.110 \\
Low-income EMs & Two-way + crisis & 0.083 & 0.042 & 0.041 & -0.074 & -0.074 & -0.102 \\
\bottomrule
\end{tabular}
\begin{flushleft}
\footnotesize \textit{Notes:} Authors' elaboration based on World Bank data. \\ The table reports LP-IV coefficients for the financial shock. The baseline specification clusters standard errors by country. The robust specification clusters standard errors by country and year and includes global crisis dummies for 1998, 2009, and 2020. The dependent variable is the cyclical component of inventory changes.
\end{flushleft}
\end{table}

\subsection{Long-run productivity and financial shock estimation under different control specifications}
\label{app:specifications_1}

Overall, the results are robust to the choice of lag length. Across country groups, the main qualitative patterns of the inventory response to a long-run productivity shock are preserved. While some differences arise in the magnitude and statistical significance of individual coefficients, the main economic conclusions remain unchanged.

For advanced economies, the positive inventory response is similar across the one-lag and two-lag specifications. In both cases, inventories increase on impact, decline gradually over the following horizons, and converge toward values close to zero at longer horizons. The response remains positive during the first three horizons, although statistical significance is somewhat stronger in the baseline two-lag specification.

Upper-middle and high-income emerging economies continue to display a temporary positive response at short horizons under both specifications. The main difference arises in the medium term where, under the one-lag specification, the response moves close to zero and then becomes positive again at longer horizons, whereas under the two-lag specification the response turns negative around horizon 3 before returning toward zero. Despite this difference, both specifications indicate that the initial inventory expansion following a long-run productivity shock is not persistent.

Lower-middle-income emerging economies also show broadly similar dynamics across lag structures. In both specifications, inventories increase on impact and in the following horizon, then decline gradually and turn negative at medium horizons before converging back toward zero by the end of the response window. The timing and shape of the adjustment are therefore largely preserved when the lag structure is reduced from two lags to one.

\begin{figure}[H]
    \centering
    \caption{Inventory Response to long-run productivity shock under alternative lag specifications}
    \label{fig:irf_productivity_1_2_lags}
    \includegraphics[width=.9\textwidth]{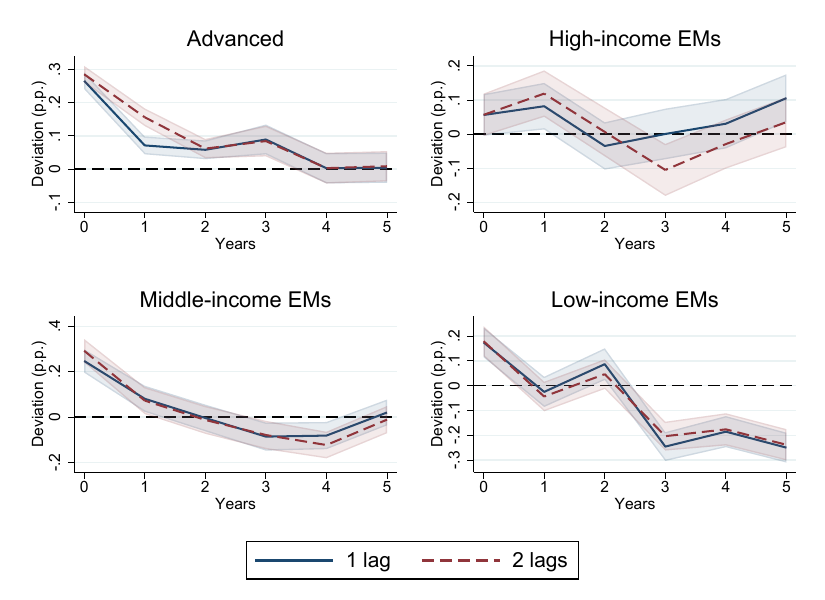}
        \begin{tabular}
    {p{15.0cm} p{width}}
    {\setstretch{1.1}
    \scriptsize
    \textit{Note}: Authors' elaboration based on World Bank data. Shaded areas represent 68\% confidence intervals.}
\end{tabular}
\end{figure}

Low-income emerging economies exhibit the most persistent negative adjustment. In both specifications, inventories respond positively on impact, decline in the following horizon, partially recover around horizon 2, and then become negative from horizon 3 onward. Although the magnitude of the responses differs somewhat across specifications, the timing of the reversal and the persistence of the contraction remain unchanged (Figure \ref{fig:irf_productivity_1_2_lags}).

Taken together, these results suggest that the estimated inventory responses to long-run productivity shocks are not driven by a particular lag-length specification. The key findings regarding the sign, timing, and persistence of inventory adjustments remain broadly robust under the alternative one-lag local-projection specification.

In the case of financial shocks, the results are robust to the choice of lag length. The estimated responses display similar signs, magnitudes, and persistence under both the one--lag and two--lag specifications, suggesting that the main findings are not driven by a particular dynamic specification. Although some differences emerge in the magnitude and statistical significance of individual coefficients, the main economic conclusions remain unchanged.

For advanced economies, changes in inventories decline in horizon 1, where the response is negative and statistically significant in both models. Changes in inventories remain significantly below zero at horizon 2 before turning positive and statistically significant at horizon 3. After this horizon, responses are no longer statistically different from zero. The magnitude and timing of the adjustment are nearly identical across lag structures.

Upper-middle and high-income emerging economies display a similar pattern under both specifications. The response becomes significantly negative at
horizon 1, indicating an immediate inventory contraction following a financial shock. This is followed by a significant rebound at horizons 2, 3 and 4, suggesting a temporary recovery of inventories after the initial adjustment. Although the two--lag specification produces smaller coefficients at some horizons, the overall dynamic pattern remains unchanged.

Lower-middle-income economies also exhibit highly stable responses across specifications. Inventories fall significantly at horizon 1, recover and become significantly positive at horizons 2 and 3, and subsequently return to values that are not statistically different from zero. The magnitude of the responses differs only marginally between the one--lag and two--lag models.

Low-income emerging economies present the most persistent inventory contraction. While the impact response is not statistically significant, inventories become significantly negative from horizons 3 to 5 under both specifications. The magnitude of the decline is larger under the two-lag model, but the timing and persistence of the adjustment remain identical. This suggests that financial shocks generate a prolonged reduction in inventories.

\begin{figure}[H]
    \centering
    \caption{Inventory Response to financial Shock under alternative lag specifications}
    \label{fig:irf_spread_1_2_lags}
    \includegraphics[width=.9\textwidth]{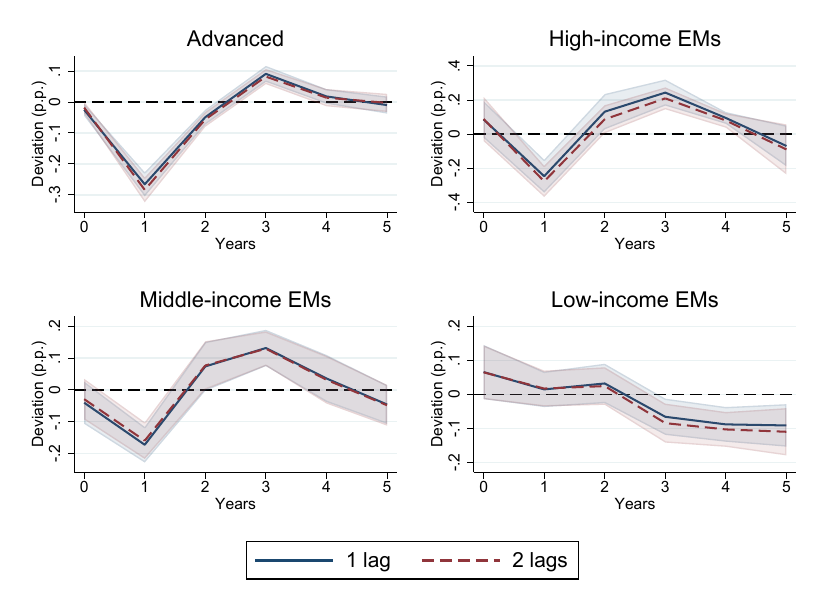}
        \begin{tabular}
    {p{15.0cm} p{width}}
    {\setstretch{1.1}
    \scriptsize
    \textit{Note}: Authors' elaboration based on World Bank data. Shaded areas represent 68\% confidence intervals.}
\end{tabular}
\end{figure}

Taken together, these results indicate that the estimated inventory responses to financial shocks are not sensitive to the choice of lag length. The key findings regarding the direction, timing, and persistence of inventory adjustments remain robust under the alternative one-lag specification in the local projection specification (Figure \ref{fig:irf_spread_1_2_lags}).

\subsection{Leave-One-Country-Out Analysis of the Productivity- and Financial-Shock Responses}\label{app:leave_one_country_out}

To assess whether individual countries drive the results, we implement a leave-one-country-out exercise. For each shock and country group, we re-estimate the local projections described in Section \ref{sec:shocks}, excluding one
country at a time. At each horizon, we compare the full-sample coefficient with the minimum and maximum estimates obtained across the re-estimations.

In general terms, the results show that the response of inventories to long-run productivity shocks is not driven by the exclusion of a single country. Although some horizons display sensitivity in magnitude, the main qualitative conclusions remain broadly unchanged.

Advanced economies exhibit a positive response during the early horizons following the productivity shock. The contemporaneous impact is positive and remains above zero through horizon 3, before converging toward values close to zero at longer horizons. The leave-one-country-out estimates show that this initial expansion is robust to the exclusion of individual countries, since the range of estimates remains positive in the early horizons. Greater sensitivity appears only at longer horizons, where the response approaches zero.

In high-income and upper-middle-income emerging economies, the response is more sensitive to the exclusion of individual countries. The baseline response is positive on impact and in horizon 1, close to zero around horizon 2, and positive again at later horizons. However, the leave-one-country-out range becomes wider from horizon 2 onward and includes both positive and negative values in several medium-term horizons. This indicates that the dynamics in this group are less robust than in advanced economies and depend more strongly on sample composition.

Lower-middle-income emerging economies show a robust positive initial response. Both the contemporaneous impact and the response in the following year remain positive across the leave-one-country-out exercises. After horizon 2, the response becomes weaker and more sensitive, with estimates moving around zero before recovering at the end of the response window. Thus, the initial accumulation of inventories is robust, while the medium-term dynamics are less stable.

\begin{figure}[H]
    \centering
    \caption{Inventory Response to long-run productivity shock leave one country out for estimation}
    \label{fig:leave_one_productivity}
    \includegraphics[width=.9\textwidth]{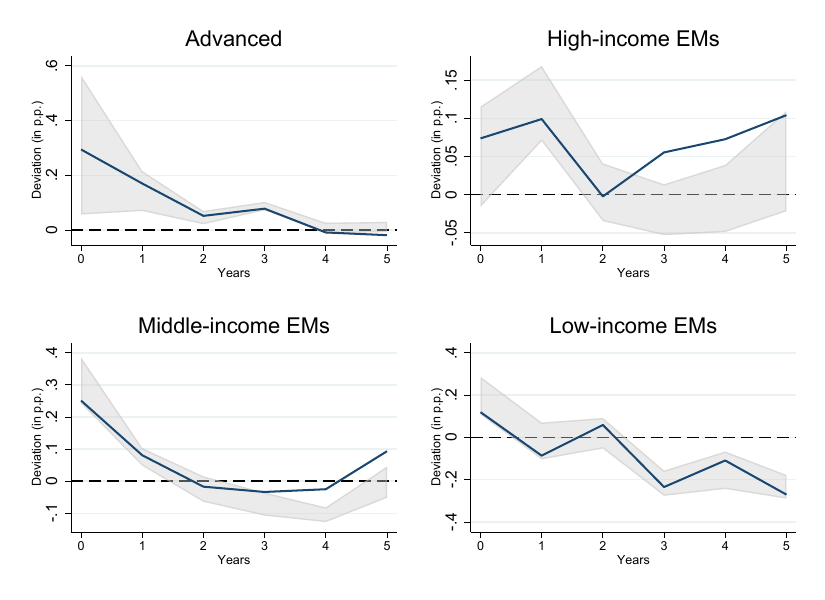}
        \begin{tabular}
    {p{15.0cm} p{width}}
    {\setstretch{1.1}
    \scriptsize
    \textit{Note}: Authors' elaboration based on World Bank data.}
\end{tabular}
\end{figure}

Low-income emerging economies display a positive contemporaneous response, followed by a more volatile adjustment. The response becomes negative from horizon 3 onward, and the leave-one-country-out range remains below zero in the later horizons. This suggests that the delayed contraction of inventories in this group is not driven by a single country. Overall, the exercise confirms that the main negative adjustment observed in low-income emerging economies is robust to changes in sample composition (Figure \ref{fig:leave_one_productivity}).

Regarding the responses to financial shocks, the results show that the estimated dynamics are generally robust to the exclusion of individual countries. In no group does the leave-one-country-out exercise produce a substantial change in the qualitative shape of the impulse-response functions, suggesting that the main results are not driven by a single country in the sample.

Advanced economies display an initial contraction in inventories following the financial shock. The response is negative on impact and reaches its largest decline in horizon 1. Thereafter, inventories recover gradually, moving close to zero around horizon 2 and becoming positive in horizons 3 and 4, before converging again toward zero at the end of the response window. The leave-one-country-out estimates show limited dispersion across horizons, indicating that both the initial contraction and the subsequent recovery are robust to the exclusion of individual countries.

In high and upper-middle-income emerging economies, the response also shows a clear short-run contraction followed by a recovery. Inventories decline in horizon 1, then turn positive from horizon 2 onward and reach their largest positive response around horizon 3. The response gradually fades afterward and becomes slightly negative by horizon 5. The leave-one-country-out range is relatively narrow in most horizons, particularly during the contraction and recovery phases, suggesting that the estimated dynamics are stable across alternative country compositions.

Lower-middle-income emerging economies follow a similar pattern. Inventories decline on impact and more strongly in horizon 1, before turning positive in horizons 2 and 3. The response then weakens and converges back toward zero in the final horizons. The leave-one-country-out estimates remain close to the baseline path, especially around the initial contraction and the subsequent recovery, indicating that the main adjustment pattern is not driven by any single country.

\begin{figure}[H]
    \centering
    \caption{Inventory Response to financial shock leave one country out for estimation}
    \label{fig:leave_one_financial}
    \includegraphics[width=.9\textwidth]{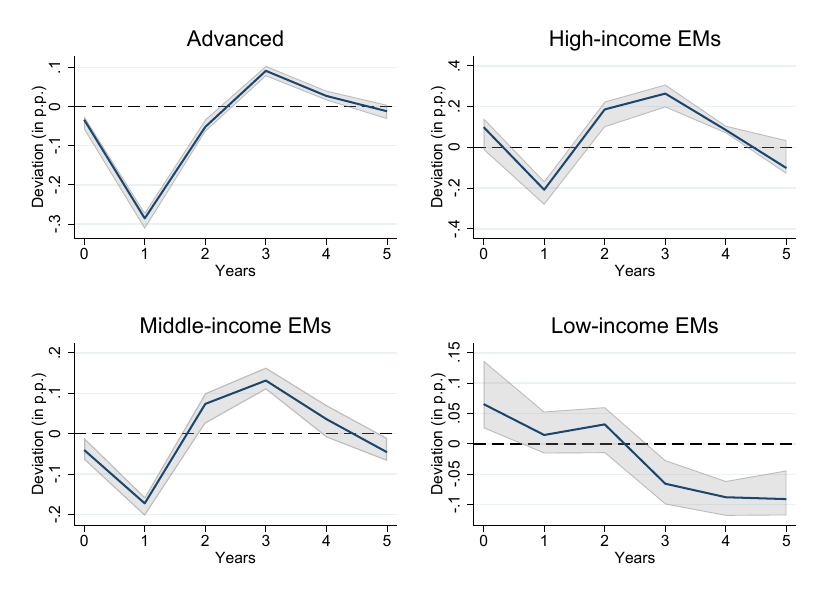}
        \begin{tabular}
    {p{15.0cm} p{width}}
    {\setstretch{1.1}
    \scriptsize
    \textit{Note}: Authors' elaboration based on World Bank data.}
\end{tabular}
\end{figure}

Low-income emerging economies display a different dynamic. Inventories increase slightly in the first horizons after the financial shock, but the response turns negative from horizon 3 onward and remains below zero until the end of the response window. The leave-one-country-out range is wider in the early horizons, suggesting greater sensitivity to sample composition at the beginning of the adjustment process. However, from horizon 3 onward, the negative response is preserved across re-estimations, indicating that the delayed contraction in inventories is a robust feature of this group (Figure \ref{fig:leave_one_financial}).

In summary, the leave-one-country-out exercises indicate that the estimated responses to long-run productivity and financial shocks are not driven by individual countries. The differences observed across re-estimations mainly affect the magnitude of some coefficients, but do not alter the central conclusions of the analysis. In particular, the initial accumulation of inventories in advanced and lower-middle-income emerging economies, as well as the delayed contraction observed in emerging low-income economies, remain present in most of the specifications considered.

\subsection{Cross-Sectional Dependence in the Local Projections}
\label{app:csd_tests}

Because the panel includes countries exposed to common global disturbances, we evaluate residual cross-sectional dependence in the local-projection specifications using Pesaran's CD test. Tables \ref{tab:pesaran_cd_productivity} and \ref{tab:pesaran_cd_financial} report the CD statistics by country group and projection horizon for the productivity and financial shocks, respectively.

For the productivity-shock specifications, which include both country and year fixed effects, the evidence of residual cross-sectional dependence is limited. The null is never rejected for low-income EMDEs, while rejections occur only at selected horizons for advanced economies and high- and upper-middle-income EMDEs. The strongest evidence appears among lower-middle-income EMDEs, where the null is rejected in most horizons.

For the financial-shock specifications, residual cross-sectional dependence is more pronounced. The null is rejected at all horizons for advanced economies, high- and upper-middle-income EMDEs, and lower-middle-income EMDEs, while it is not rejected for low-income EMDEs. This pattern is consistent with the global nature of the financial shock and with stronger exposure of more financially integrated economies to common global financial conditions. The results indicate that instrument relevance is moderate across groups and horizons. The weakest values are observed among low-income EMDEs, although the effective $F$-statistics generally remain close to or above the 20 percent maximal relative bias threshold. Since the estimated financial-shock responses for this group are small and imprecise, the main conclusions of the paper do not rely on strong financial-shock effects in low-income economies.

These results do not invalidate the local-projection estimates, but they motivate the robustness exercises based on alternative inference schemes, including year-clustered standard errors and two-way clustering by country and year. The main qualitative patterns are preserved under these alternative specifications.

\begin{table}[H]
\centering
\caption{Pesaran CD Tests for Cross-Sectional Dependence: Productivity Shock}
\label{tab:pesaran_cd_productivity}
\small
\begin{tabular}{lcccccc}
\toprule
\textbf{Group} & \textbf{$h=0$} & \textbf{$h=1$} & \textbf{$h=2$} & \textbf{$h=3$} & \textbf{$h=4$} & \textbf{$h=5$} \\
\midrule
Advanced        & -1.08 & -2.13** & -1.71* & -1.72* & -1.50 & -2.08** \\
High-income EMs      & -2.32** & -2.20** & -1.83* & -2.10** & -1.57 & -1.44 \\
Middle-income EMs         & -2.34** & -2.08** & -2.18** & -2.35** & -2.06** & -1.80* \\
Low-income EMs          & -1.51 & 0.01 & -0.29 & -0.34 & -0.89 & -1.03 \\
\bottomrule
\end{tabular}
\begin{flushleft}
\footnotesize \textit{Notes:} Authors' elaboration based on World Bank data.\\ The table reports Pesaran CD statistics computed from the residuals of the productivity-shock local projections. The null hypothesis is weak cross-sectional dependence. The specifications include country and year fixed effects, as well as two lags of GDP growth and the inventory ratio. $^{***}$, $^{**}$, and $^{*}$ denote rejection at the 1, 5, and 10 percent levels, respectively.
\end{flushleft}
\end{table}

\begin{table}[H]
\centering
\caption{Pesaran CD Tests for Cross-Sectional Dependence: Financial Shock}
\label{tab:pesaran_cd_financial}
\small
\begin{tabular}{lcccccc}
\toprule
\textbf{Group} & \textbf{$h=0$} & \textbf{$h=1$} & \textbf{$h=2$} & \textbf{$h=3$} & \textbf{$h=4$} & \textbf{$h=5$} \\
\midrule
Advanced        & 20.21*** & 19.68*** & 21.80*** & 21.09*** & 21.97*** & 22.38*** \\
High-income EMs       & 8.90***  & 7.40***  & 9.39***  & 11.22*** & 8.35***  & 8.70***  \\
Middle-income EMs         & 5.95***  & 4.71***  & 4.63***  & 5.72***  & 5.33***  & 5.64***  \\
Low-income EMs          & -0.99    & -1.05    & -0.19    & 0.18     & 0.24     & 0.58     \\
\bottomrule
\end{tabular}
\begin{flushleft}
\footnotesize \textit{Notes:} Authors' elaboration based on World Bank data.\\ The table reports Pesaran CD statistics computed from the residuals of the LP-IV estimations of the financial shock. The null hypothesis is weak cross-sectional dependence. The specifications include country fixed effects and two lags of GDP growth and the inventory-cycle variable. We instrument the financial shock using Federal Reserve monetary surprises and changes in global macro-financial uncertainty. $^{***}$ denotes rejection at the 1 percent level.
\end{flushleft}
\end{table}

\newpage

\section{Responses of Other Macroeconomic Variables to Productivity and Financial Shocks} \label{app:others_variables}
\renewcommand{\thetable}{G\arabic{table}}
\renewcommand{\thefigure}{G\arabic{figure}}

\setcounter{table}{0}
\setcounter{figure}{0}

This appendix presents the responses of GDP, consumption, exports, and imports to long--run productivity and financial shocks across different groups of countries. These impulse responses help to characterize inventory dynamics following each type of shock across country groups.

We estimate the responses using the local-projection methodology and the same specification as in the main text. For each horizon, we estimate a separate regression with country fixed effects and lagged controls, using standard errors clustered at the country level. For long-run productivity shocks, we also use a robust estimator to reduce the influence of outliers.

\begin{figure}[H]
    \centering
    \caption{Responses of Macroeconomic Aggregates to a Long-Run Productivity Shock, by Country Group}
    \label{fig:irf_TFP_others}
    \includegraphics[width=.9\textwidth]{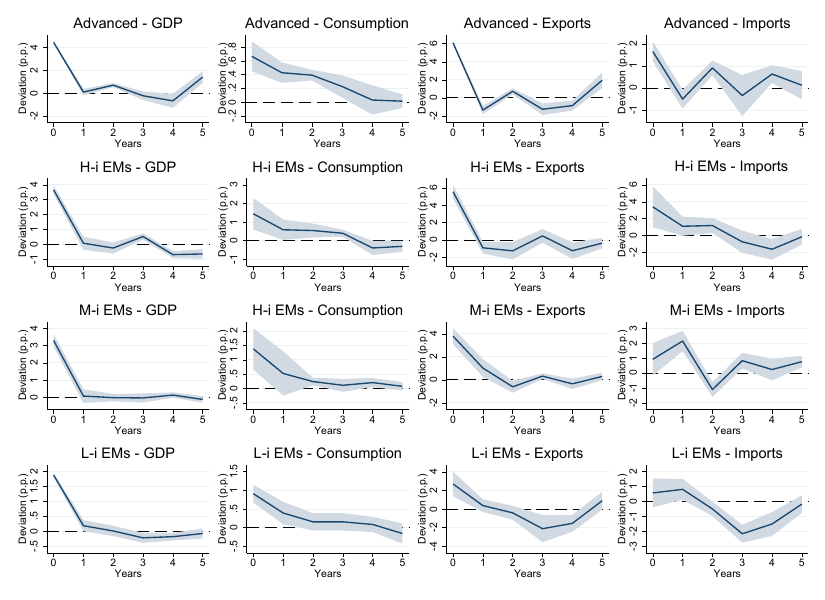}
        \begin{tabular}{p{15.0cm} p{width}}
    {\setstretch{1.1}
    \scriptsize
    \textit{Note}: Authors' elaboration based on World Bank data. Shaded areas represent 68\% confidence intervals.}
\end{tabular}
\end{figure}

Figure \ref{fig:irf_TFP_others} shows that a positive long-run productivity shock produces an immediate expansion in GDP, consumption, exports, and imports across country groups. However, the estimated paths differ in their persistence and stability across country groups. In advanced economies, output rises on impact and then gradually returns toward values close to zero, while consumption adjusts more smoothly and external trade displays short-run fluctuations before stabilizing. In upper-middle-income emerging economies, the initial expansion is followed by a relatively rapid decline, particularly in GDP and trade flows, with some responses turning negative at medium horizons. In emerging middle-income economies, the point estimates for GDP and consumption return toward zero relatively quickly, while imports and exports display larger oscillations. Finally, emerging low-income economies exhibit point-estimate paths with large oscillations, with output and consumption losing strength rapidly and external trade responses turning negative in several horizons.

Overall, the figure suggests that productivity-related shocks generate short-run gains in aggregate activity, while the estimated responses tend to fade sooner and fluctuate more at lower income levels. This pattern is consistent with the presence of stronger financial, external, and productive frictions, which may limit the persistence of productivity gains and generate sharper reversals in aggregate demand components.

\begin{figure}[H]
    \centering
    \caption{Responses of Inventory Changes to a Financial Shock, by Country Group (aggregate demand components)}
    \label{fig:irf_spread_others_vars}
    \includegraphics[width=.9\textwidth]{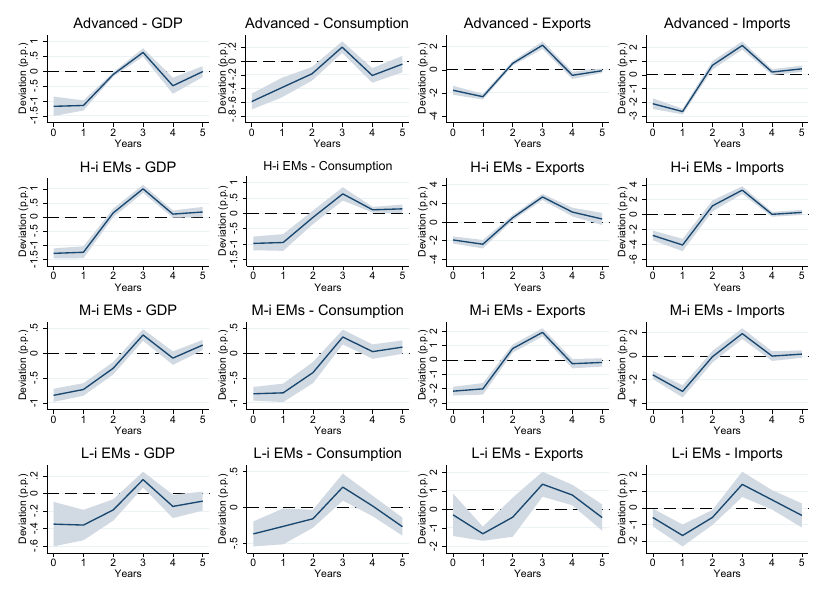}
        \begin{tabular}{p{15.0cm} p{width}}
    {\setstretch{1.1}
    \scriptsize
    \textit{Note}: Authors' elaboration based on World Bank data. Shaded areas represent 68\% confidence intervals.}
\end{tabular}
\end{figure}

Figure \ref{fig:irf_spread_others_vars} shows that a tightening in global financial conditions generates an initial contraction in GDP, consumption, exports, and imports across country groups. The negative response is concentrated in the first horizons, suggesting that financial shocks operate through weaker aggregate demand, lower external trade, and tighter financing conditions. After this initial decline, most variables display a partial recovery, with responses turning positive around the third year, especially in advanced and emerging economies.

The adjustment pattern differs across levels of development. In advanced and emerging high-income economies, the estimated responses follow more regular paths, which is consistent with a clearer transmission of global financial conditions to domestic activity and trade. In lower-middle and low-income emerging economies, the responses are more volatile and less precisely estimated, as reflected in wider confidence bands. Overall, the figure suggests that financial shocks generate short-run macroeconomic contractions, while differences in the subsequent recovery may be related to countries' financial depth, external exposure, and capacity to absorb global financing disturbances.

\end{document}